\documentstyle[12pt]{article}
\newcommand{\half}{\mbox{$\frac{1}{2}$}}
\begin{document}
\def\yes{y }
%
\def\hasepsf{y }
\ifx\hasepsf\yes\message{Including pictures!}
        \input{epsf.sty}
\else
        \newcommand{\epsffile}[1]{}
        \newlength{\epsfxsize}
        \newlength{\epsfysize}
\fi
\begin{titlepage}
\noindent\begin{minipage}{7cm}
{\Huge FAU}\\
{\it Friedrich Alexander Universit\"at}\\
Inst. Theor. Phys. III\\
Staudtstrasse 7\\
D-91058 Erlangen, Germany
\end{minipage}
\hspace*{\fill}\begin{minipage}{7cm}
\flushright
preprint FAU-T3-94/1\\
hep-ph/9410254
\end{minipage}
\vspace*{2 cm}

\begin{center}
{\Large \bf The kinetic energy and and the geometric structure in the
$B=2$ sector
of the Skyrme model:\\
A study using the Atiyah-Manton Ansatz}
\end{center}

\begin{center}
{\bf Niels R. Walet}
\end{center}
\vspace*{\fill}
\begin{center}
{\em Submitted to Nuclear Physics A}
\end{center}

\end{titlepage}

\title{%
The kinetic energy and and the geometric structure in the $B=2$ sector
of the Skyrme model:\\
A study using the Atiyah-Manton Ansatz}

\author{Niels R. Walet\thanks{electronic address:
nwalet@theorie3.physik.uni-erlangen.de}\\
Institut f\"ur theoretische Physik III, Universit\"at Erlangen-N\"urnberg,\\
D-91058 Erlangen, Germany
}

\date{\today}
\maketitle

\begin{abstract}
We study the construction of the collective-coordinate manifold in the
baryon number two sector of the Skyrme
model. To that end we use techniques of adiabatic large amplitude
collective motion, which treat potential and kinetic energy on an equal
footing. In this paper the
starting point is the Ansatz proposed by Atiyah and Manton
(Phys.~Lett.~{\bf 438B}, 222 (1989)), which allows a study
of the dynamics using a finite and small number of
variables. From these variables we choose a subset of collective ones.
We then study the behavior of inertial parameters along parts
of the collective manifold, and study the dynamical parts of the interaction.
\end{abstract}

\section{Introduction}
In this paper we study an attempt to model the nuclear force from an
underlying effective field theory, that mimics some aspects
of the ``correct'' theory of strong interactions, namely QCD.
This is in sharp contrast to the standard effective models
of the nuclear interaction, which have been derived mainly through
a description in terms of (fictitious) meson exchanges. In general they
have received remarkably little direct input from QCD
(see Ref.~\cite{Bonnpot} for a discussion of several approaches).
Nevertheless such an approach is both a quantitative and qualitative success.

Another possible approach, which has only been pursued in recent years,
is the use of models inspired by the large $N_c$ limit of QCD
\cite{tHooft,Witten}.
The prime example is the Skyrme model \cite{Skyrme62}, a non-linear
theory of interacting pions, where baryons appear as topological defects -
and this implements confinement.
The model was conceived and forgotten in the early sixties, and
was revived only in the early eighties by Witten and collaborators \cite{ANW}.
It has since been used successfully to describe the properties of baryons.
Of course
it was soon realized that one could use the model to describe baryon-baryon
interactions as well \cite{VinhMau,JacksonJP85}. Actually even this idea seems
to have been discovered by Skyrme \cite{PerringSkyrme62}.

The baryon-baryon interaction derived in this model was plagued by a lack of
central attraction. In the early works the interaction was derived using
the product approximation, which assumes that the two Skyrmions do not
deform as they approach. Furthermore one ignored the mixing between
the nucleon and its excited states, which gives an attractive contribution
to the central force.
This question has recently been  investigated  in some detail, and it was
found that the reintroduction of a finite number of colors ($N_c$),
 together with the use of a Born-Oppenheimer
approximation is necessary to describe this part correctly \cite{WaletAH92,%
WaletAmado93}.

Apart from the static interaction (i.e., the part of the energy
independent of velocities)
there is a dynamical structure of the Hamiltonian as described by
the kinetic energy. The inertial parameters are in general position dependent.
If we use the fact that the mass tensor can be used as
a natural metric on the configuration manifold, this shows that
that coordinate space is curved. Actually little is know about these
parameters. The only calculations
have been performed using the product ansatz, and are thus only
good for large separations. One of the important parts of the $NN$ force,
the spin-orbit interaction, derives purely from these kinetic terms.
One finds that, even when the separation is large, and the product
Ansatz is supposed to work,
the sign of the spin-orbit force \cite{RiskaDannbom88,Otofuji,%
AmadoSW93,AmadoSW94,Abada94} does not agree with phenomenology.
This problem  has not yet been resolved.

The potential and kinetic energies can be derived from a knowledge
of the solutions of the Skyrme model for finite separations.
Unfortunately no such solutions exist. The best calculations up to date
\cite{WalhoutWambach91} calculate a set of fields obtained by
imposing a constraint that is chosen in advance, and
not determined self-consistently. As an alternative there exist the
set of parametrized fields introduced by Atiyah and Manton
\cite{AtiyahManton89}. These authors use an interesting technique to obtain
a field of given topological quantum number from a parametrized
instanton field. Its advantage is that the instantons, which are specified
by a finite number of parameters, induce an effective Lagrangian for the
Skyrme model depending on the instanton parameters.
One can attempt to choose combinations among those that describe the
collective manifold. As a first step towards this goal a calculation,
 similar in nature to that
of Walhout and Wambach, of the static potential was performed
by Hosaka {\em et al} \cite{HosakaOA91}. This gave reasonable answers.

Since one can perform exact numerical solutions for the same
problem, the greater promise of this Ansatz lies in the understanding
of the geometrical structure of the collective manifold, which
is closely tied to the kinetic energy.
Atiyah and Manton have recently \cite{AtiyahManton93} analyzed some
aspects of this problem analytically. Several of the more
quantitative aspects of this approach require numerical calculations,
and cannot be resolved by analytical means.
This paper is an attempt to study
such an  approach. There is a related idea to use the Atiyah-Manton Ansatz
 to study peripheral scattering \cite{Gisiger94}. Our work has no overlap
with this approach, since we shall concentrate on short and intermediate
distances. We believe that the long distance behavior can be best described
by the product approximation, maybe suitably symmetrized.

The determination of collective coordinates
in classical mechanics, or -- equivalently -- time-dependent mean field
theories, has been studied extensively over the past 20 years
(see Ref.~\cite{KleinWD91} for a discussion).
As discussed in Ref.~\cite{KleinWD91} the theory of adiabatic
large amplitude collective motion (ALACM) is developed well enough
to construct algorithms that are singularly well suited for this task.
In this note we shall apply the AM Ansatz, and mix it with the
notions of ALACM in
order to study the structure of the adiabatic manifold. Similar
calculations can be -- and should be -- performed by putting the Skyrme
model on a grid. Since such a task is much more complex, both numerically
and analytically, it is much harder to gain  insight in  that way.

The paper is organized as follows. In Sec.~\ref{sec:SmAM} we succinctly
introduce the Skyrme model as well as the product Ansatz and the
Atiyah Manton Ansatz. In Sec.~\ref{sec:LACM} we give a short introduction
to the theory of Large Amplitude Collective Motion, as far
as relevant for the current paper. In Sec.~\ref{sec:num} we introduce
the numerical techniques used in our calculations. In
Sec.~\ref{sec:results} we discuss the results of these calculations.
Finally, in Sec.~\ref{sec:disc}, we give a discussion and outlook.
In three appendices we discuss which classes of solutions have a reflection
symmetry,
we give a detailed analysis of the fluctuations around the  $B=2$ hedgehog and
we discuss the algorithm used in obtaining self-consistent solutions
for the LACM equations.

\section{Skyrme model and Atiyah-Manton Ansatz \label{sec:SmAM}}
\subsection{The model}
The ``standard'' Skyrme model is based on the non-linear sigma model,
extended by a quartic interaction term and a pion mass term.
The Atiyah-Manton Ansatz can only be used in the (chiral) limit
of zero pion mass, where the model is defined by the Lagrange density
\begin{equation}
{\cal L}=\frac{f^{2}_{\pi}}{4}Tr(\partial_{\mu}U\partial^{\mu}U^{\dagger})
+\frac{1}{32g^{2}_{\rho}}
Tr[U^{\dagger}\partial_{\mu}U,U^{\dagger}\partial_{\nu}U]
[U^{\dagger}\partial^{\mu}U,U^{\dagger}\partial^{\nu}U],
\end{equation}
where $U$ is a unitary two-by-two matrix-valued field satisfying the
boundary condition $U=1$ at infinity. As has been discussed many times
before, this model has a topologically conserved quantum current.
The charge of this current is
identified with baryon number $B$. Finally, one can also separate the $U$
field in a $\sigma$ and pion field,
\begin{equation}
U = \frac{1}{f_\pi}(\sigma + i{\bf\pi}\cdot{\bf\tau}).
\end{equation}
On occasion it may be useful to look at the vector pion field.

If we rescale the units of time and length, $x\rightarrow x/(gf_\pi)$,
(the so-called Skyrme units),
the Lagrange density takes on the slightly more convenient form
\begin{equation}
{\cal L}=\frac{f_{\pi}}{g_\rho}
\left(\frac{1}{4}Tr(\partial_{\mu}U\partial^{\mu}U^{\dagger})
+\frac{1}{32}
Tr[U^{\dagger}\partial_{\mu}U,U^{\dagger}\partial_{\nu}U]
[U^{\dagger}\partial^{\mu}U,U^{\dagger}\partial^{\nu}U]\right),
\end{equation}
 where ${f_{\pi}}/{g_\rho}$ is the Skyrme unit of energy.%
\footnote{There are many inequivalent definitions of the Skyrme units in the
literature, depending on the definition of $f_\pi$. In our definition
the S.U. of energy is equivalent  to 11.18 MeV
(for the ``standard'' values of $f_\pi=54.1~\rm MeV$ and $g=4.84$),
 and the S.U. of length is 0.754 fm.}
Finally the Skyrme Lagrangian can easily be reformulated in terms of the
Sugawara variables (Lie-algebra valued currents) $L$,
\begin{equation}
L_\mu = U^{-1}\partial_\mu U = i\,l_\mu^a \tau_a,
\label{eq:defLmu}
\end{equation}
and we have
\begin{equation}
{\cal L}=
\frac{1}{2}l_\mu^a l^\mu_a
+\frac{1}{4}\left[(l_\mu^a l^{\mu a})^2-l_\mu^a l_\nu^al^{\mu b} l^{\nu b}
\right]
\end{equation}

\subsection{Structure of the $B=2$ manifold}
Preliminary investigations of the structure of the $B=2$ manifold
have been performed in great detail using the product Ansatz,
and are discussed in the reviews \cite{BrownZahed86,NymanRiska90}.

Let us recapitulate the necessary details. For baryon number one
the basic solution is the hedgehog,
\begin{equation}
U = \exp(i{\bf\tau} \cdot \hat r f(r)),
\end{equation}
with $f(0) = \pi$ and $f(\infty) = 0$.
The baryon density has spherical symmetry, and the pion field
points radially outward at each point of this surface.
Actually one can perform a constant isorotation on the quantity $U$,
$AUA^\dagger$, without changing the energy. This changes the direction
of the pion field, and thus ``grooms'' the hedgehog.

One can use these solutions  to construct the so-called
 product Ansatz. This uses  the fact that the product of
two U fields each with baryon number one, with arbitrary centers and grooming,
has baryon number two.
As Skyrme already understood, this Ansatz is only good at large separations.
At smaller separations the two $U$ fields no longer commute, and an asymmetry
between the treatment of the two hedgehogs appears. Furthermore, the bound
state in the $B=2$ sector of the model, named  ``donut'' after the toroidal
symmetry of the baryon density
\cite{VWWW87,Verbaarschot87,Manton88a,KopeliovicStern87},
 can not be described by the product Ansatz.

\begin{figure}
\epsfysize=12cm
\centerline{\epsffile{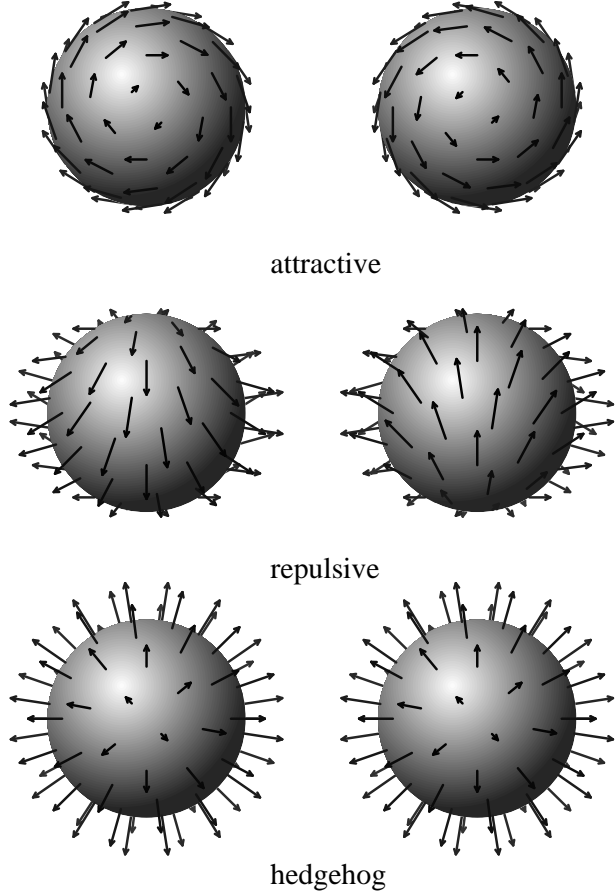}}
\caption{A schematic representation of the three channels,
as they would appear
in the limit of large separations. The spheres represent a surface of constant
baryon density, and the arrows show the pion field on this surface. Here we
have identified space and isotopic space.
\label{fig:channels}}
\end{figure}

One way out is to solve the Skyrme model numerically on a grid, with
imposition of the separation of the two Skyrmions. This approach has been taken
by Wambach and his collaborators \cite{WalhoutWambach91,WalhoutWambach92}.
Actually they never solve for all possible
states of two Skyrmions, but limit themselves to a subset of configurations
with definite symmetries under reflections, called the attractive, repulsive
and hedgehog channels. These configurations are represented schematically,
using the product Ansatz solutions in which they go over for large
separations,  in
Fig.~\ref{fig:channels}. (A similar study using the Atiyah-Manton Ansatz
was performed in Ref.~\cite{HosakaOA91}.)

The three channels are first of all the hedgehog channel,
where we perform no rotation
on the two hedgehogs. All states have the same reflection symmetries as a
single
baryon-number two hedgehog.
Second is the repulsive channel, where we perform a $90^\circ$ rotation
of each hedgehog ($180^\circ$ relative angle) around the axis through the
centers of the hedgehog. This is called repulsive since there is a strong
repulsion in this channel. Finally there is the attractive channel where
each hedgehog is rotated by $90^\circ$ degrees about an axis orthogonal
to the line connecting the two centers.

{}From the caricatures of the pion field drawn here one can easily see
that the fields have a certain symmetry under reflection in each of the
three coordinate planes.
These symmetries are discussed in greater detail in Appendix \ref{app:symm}.
The idea is that these symmetries endow a special importance to these
configurations, since they are at least stationary against small
fluctuations breaking the symmetries.

\subsection{Atiyah-Manton Ansatz}

As discussed in \cite{AtiyahManton93} one can derive a Skyrme field
from an instanton
field by integrating the time component of the gauge potential,
\begin{equation}
U(\vec{x}) = C {\cal S} \left\{P \exp\left[\int_{-\infty}^\infty
-A_4(\vec{x},t) dt \right]\right\}
C^\dagger .
\label{eq:holonomy}
\end{equation}
Here ${\cal S}$ is a constant matrix, chosen such that  $U$ decays
to $1$ at infinity, and $C$ describes an overall grooming. For the current
work, where we shall only consider $C$ near the identity, it is convenient
to parametrize
\begin{equation}
C = \exp(i\vec \tau \cdot \vec \theta). \label{eq:deftheta}
\end{equation}
For the Jackiw-Nohl-Rebbi (JNR) instanton of charge $k=2$ we have
\cite{JNR}
\begin{equation}
A_4(\vec{x},t) = \frac{i}{2} \frac{\vec{\nabla}\rho}{\rho}\cdot\vec{\tau},
\;\;\;\rho=\sum_{l=1}^{k+1} \frac{\lambda_l}{|x-X_l|},
\end{equation}
and we should use ${\cal S}=-I$, to obtain a field of baryon number $B=2$.

In order to solve for the AM value of $U$ we convert the integral
(\ref{eq:holonomy}) to the solution of a differential equation. First introduce
\begin{equation}
\tilde{U}(\vec{x},\tau) = C{\cal S} \left\{
P \exp\int_{-\infty}^\tau -A_4(\vec{x},t)dt\right\}C^\dagger
{}.
\end{equation}
This function satisfies the differential equation
\begin{equation}
\partial_\tau \tilde{U}(\vec{x},\tau)
= -\tilde{U}(\vec{x},\tau)A_4(\vec{x},\tau),
\end{equation}
with initial condition $U(\vec{x},-\infty)={\cal S}$. The function
$U(\vec{x})$ is obtained as the limit for $\tau \rightarrow\infty$ of
$\tilde U$.
Actually we prefer to work with the Sugawara variables $L_\mu$,
Eq.~(\ref{eq:defLmu}), so that
the Lagrange density does not contain explicit derivatives.
These can also be calculated as the large $\tau$ limit of a quantity
$\tilde L_\mu$, which satisfies the differential equation
\begin{equation}
\partial_\tau \tilde L_\mu(\vec{x},\tau) =
[A_4(\vec{x},\tau),\tilde L_\mu(\vec{x},\tau)]
- \partial_\mu A_4(\vec{x},\tau).
\label{eq:Ltmu}
\end{equation}
Here the boundary condition is $\tilde L_\mu(\vec{x},-\infty)=0$.
By differentiating the  differential equations (\ref{eq:Ltmu})
we can obtain expressions for  derivatives of $L_\mu$ with respect to the
instanton parameters $\lambda_l$ and $X_l$, which will be needed later.

The field $U$ defined in Eq. (\ref{eq:holonomy})
does not have an explicit time dependence.
There is an
implicit dependence, due to a possible variation of the
instanton parameters as well as the unitary matrix $C$ with  time.
Let us denote the parameters $\{\lambda_l,X_l,\vec\theta\}$
collectively by $\xi$. We then have
\begin{equation}
L_0 = {\dot \xi}^\alpha U^\dagger \partial_{\xi^\alpha} U
\equiv
 {\dot \xi}^\alpha L_{,\alpha}.
\label{eq:Leff}
\end{equation}
If we substitute this in the Lagrangian we obtain the form
\begin{equation}
L = T - V,
\end{equation}
with
\begin{equation}
V = \half \sum_{i,a}l^a_i l^a_i + \mbox{$\frac{1}{4}$}
\left[\left(\sum_{i,a}l^a_i l^a_i\right)^2 -\sum_{ij}
\left(\sum_{a}l^a_i l^a_j\right)^2\right],
\end{equation}
and
\begin{eqnarray}
T &= & \half \dot \xi ^\alpha \dot B_{\alpha\beta}\xi ^\beta ,\\
B_{\alpha\beta} & = &
\sum_a l_{,\alpha}^al_{,\beta}^a
+2\left[\sum_a l_{,\alpha}^al_{,\beta}^a\sum_{i,b}l^b_i l^b_i
-\sum_I\left(\sum_{a}l^a_i l^a_{,\alpha}\sum_{b}l^b_i l^b_{,\beta}\right)
\right].
\end{eqnarray}

The Lagrangian is quadratic in the time-derivatives due to the special nature
of the Skyrme model. This also means that we have broken the Lorentz invariance
of the original equations, so  that (\ref{eq:Leff}) can only be
used to describe adiabatic (small velocity) motion.

In the  Atiyah-Manton Ansatz we have thus replaced  the general matrix
$U(x)$, with an infinite number of parameters, by a
form  parametrized by 18 parameters,
$C U(x|\xi)C^\dagger$.
 (At each point we can extract at most  16  dynamical parameters,
we can always form at least 2 spurious combinations. For the channels with
additional symmetry discussed in this paper, we always find one more spurious
combination.)
Even these 15 or 16 dynamical variables are too much to describe the collective
motion. We wish to study the behavior of the slowest modes
among those appearing in (\ref{eq:Leff}). The techniques for
such an approach, in the case that the Hamiltonian is quadratic in momenta,
are well-developed \cite{KleinWD91}. In the next section we will recapitulate
one of the possible mechanisms of solution, that will prove convenient in
the current context.

\begin{figure}
\epsfxsize=8cm
\centerline{\epsffile{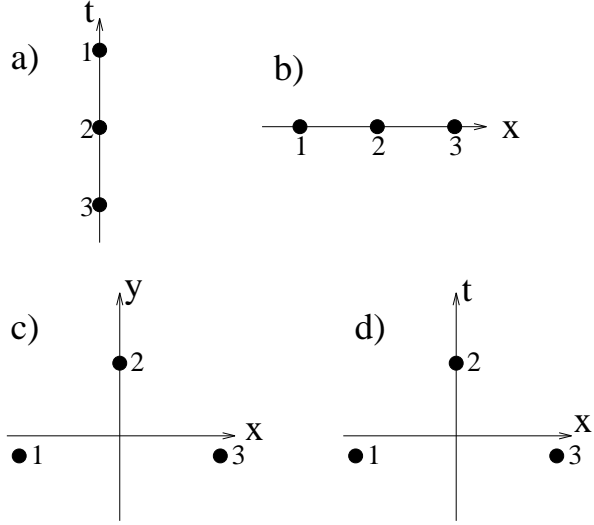}}
\caption{A schematic representations of the pole positions that describe the
three channels in the Atiyah Manton Ansatz. The pole configuration a)
represents a $B=2$ hedgehog, b) states in the hedgehog-hedgehog channel.
The triangle in c) represents the attractive states, and d) the repulsive
states.
\label{fig:poles}}
\end{figure}

Let us now enumerate those configurations, that will be relevant for the
rest of the discussion. Here we closely follow \cite{HosakaOA91},
\begin{enumerate}
\item Product Ansatz.\\
The choice  $\lambda_2\gg\lambda_1,\lambda_3$ describes a product Ansatz
solution with centers at $\vec X_1$ and $\vec X_3$.
In this limit one finds
\begin{equation}
U(\vec{x}) \approx
G_1^\dagger  U_1(\vec{x}-\vec{X}_1) G_1 G_3^\dagger U_1(\vec{x}-\vec{X}_3) G_3,
\end{equation}
where  the  iso-rotational matrices $G$ are given by
\begin{equation}
G_i = \frac{(T_i-T_2)+i(\vec{X}_i-\vec{X}_2)\cdot {\vec\tau}}
         {\sqrt{(T_i-T_2)^2 +(X_i-X_2)^2}}.
\end{equation}
\item $B=2$ hedgehog.\\
The hedgehog solution is found for the choice $\lambda_1=\lambda_3=1$,
$\lambda_2=32$, $\vec X_i=0$, $T_1=-T_3=9.5$ and $T_2=0$.
\item The hedgehog channel.\\
In general states in the hedgehog {\em channel} are described by a pole
configuration with all three poles on the $x$-axis,
$\lambda_1=\lambda_3=1$,
$\lambda_2=\lambda$, $X_{11}=-X_{31}=-X$, $X_{21}=0$,
$X_{i(2,3)}=0$, $T_i=0$. The problem is that, even though the states in the
hedgehog channel do have the same symmetry as the hedgehog, there no
continuous path from one to the other conserving the symmetry.
The only common point, where we let the three poles coincide in the origin,
corresponds to a state with $B=0$.
We shall turn to this problem in a future section, when we study the hedgehog
numerically.
\item The attractive channel.\\
Here we have
$\lambda_1=\lambda_3=1$,
$\lambda_2=\lambda$, $X_{11}=-X_{31}=-X$,
$X_{12}=X_{32}=Y$, $X_{22}$ free, all remaining components 0. Actually it was
argued in \cite{HosakaOA91} that a simple relation between the parameter
$\lambda_2$ and the shape of the triangle (see below) describes the
attractive channel best.

\item The repulsive channel.\\
Here we have the three poles aligned along the $x$-axis,
$\lambda_1=\lambda_3=1$,
$\lambda_2=\lambda$, $X_{11}=-X_{31}=-X$,
$T_{1}=T_{3}=T$, $T_{2}$ free, all remaining components 0. Again the
authors of Ref.~\cite{HosakaOA91} used the same relation as in the
attractive channel between the shape of the triangle and $\lambda$.
\end{enumerate}

Let us use the ideas of \cite{HosakaOA91} to study the configurations
in the three channels. The authors of this paper use what can be called
a turning-point approximation: At each point of the
path they only consider the potential energy
as if it were a stationary quantity. Of course this is only true if we tune
the total energy at which we work to be equal to the potential energy.
We thus end up in a turning point, since the kinetic energy has to be zero!
Furthermore
 we can then use  a virial argument to show that the quadratic
and quartic terms in the Skyrme Lagrangian contribute in equal parts
to the energy.

The formalism in the next section tries to improve on such an approach
by including information about the kinetic energy. As
as a first guess the turning-point approach is fast and cheap.
Let us look at the attractive and repulsive channel, and study the
dependence of the energy on the form and weights of the triangles.
The only remaining parameter, the size of the triangle,
 can be determined from a virial argument, since $E_2=E_4$
at a stationary point.
The hedgehog-hedgehog channel is of no special interest since it is specified
by only two parameters, one of which follows from the virial argument. The
remaining parameter then specifies the separation of the two solitons.

In order to study the energy we parametrize the triangles by introducing
a second parameter $\lambda_T$. In Ref.~\cite{HosakaOA91} it was assumed
that this parameter equals $\lambda_W$, the weight $\lambda_2$.
The parametrization takes the form
\begin{eqnarray}
\lambda_1&=&\lambda_3=1,\nonumber\\
\lambda_2 &=& \lambda_W,\nonumber\\
X_{11}&=&-D/\sqrt{(1-1/(1+\lambda_T)^2)},\nonumber\\
X_{31}&=& D/\sqrt{(1-1/(1+\lambda_T)^2)},\nonumber\\
X_{12}&=&-D/(1+\lambda_T),\nonumber\\
X_{32}&=&-D/(1+\lambda_T),\nonumber\\
X_{22}&=&D
\label{eq:tripar}
\end{eqnarray}
for the attractive channel. For the repulsive case $X_2$ should be exchanged
with  $T$.
We now study the energy as a function of
$\lambda_T$ and $\lambda_W$ for constant separation. The definition of
separation is somewhat arbitrary. We use a rotationally invariant
definition through
the size of the quadrupole moment of the baryon-number distribution
($\langle f(\vec x) \rangle \equiv \int d^3x B(\vec x) f(\vec x)$)
\begin{eqnarray}
R^4/4 &=& \langle x^2\rangle^2+\langle y^2\rangle^2+\langle z^2\rangle^2
\nonumber\\&&
-\langle x^2\rangle\langle y^2\rangle-\langle x^2\rangle\langle z^2\rangle
-\langle y^2\rangle\langle z^2\rangle
+3(\langle xy\rangle^2+\langle xz\rangle^2+\langle yz\rangle^2).
\label{eq:defR}
\end{eqnarray}
The definition in Ref.~\cite{HosakaOA91}
is based on the $11$ component of the quadrupole tensor.
As all sensible definitions should, both reduce to the
separation between the centers
of two isolated Skyrmions for large $R$.

\begin{figure}
\epsfxsize=8cm
\centerline{\epsffile{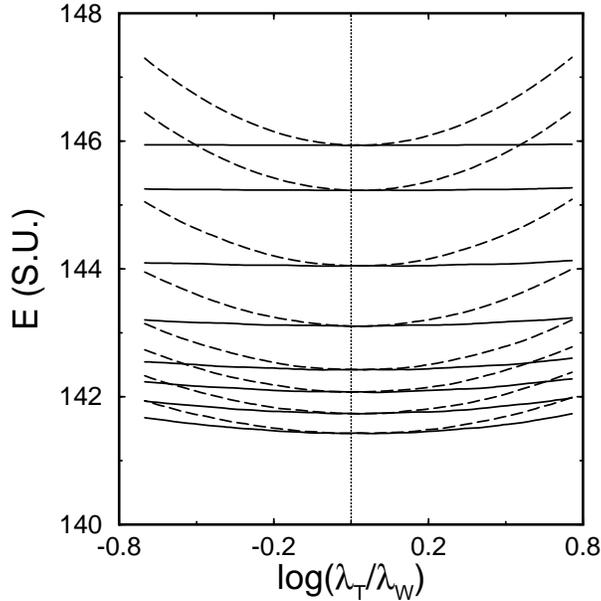}}
\caption{The energy at constant separation in the attractive channel
as a function of $\log (\lambda_W/\lambda_T)$.
{}From bottom to top the curves correspond to values of $R$
of 1.2,1.3,1.4,1.5,1.7,2.0,2.5,3.0 S.U.. Dashed lines have $\lambda_T<1$
and solid lines $\lambda_T>1$.
\label{fig:Es1}}
\end{figure}
In Fig.~\ref{fig:Es1} we show the results for the attractive channel. Here
the curves represent the energy at constant separation.
It appears to be convenient to plot them
as a function of $\log (\lambda_W/\lambda_T)$.
We find two solutions of minimal energy for  given $R$, both for
$\lambda_W=\lambda_T$.
This is due to the known inversion symmetry under $\lambda
\rightarrow 1/\lambda$, where $\lambda>1$ corresponds to separation along the
$x$-axis, and $\lambda<1$ to separation along the $y$-axis.
(There may be a hint in our calculation that slightly unequal values
of $\lambda_W$ and $\lambda_T$ may actually be the optimal solution,
but that is probably due to a numerical inaccuracy.) A careful numerical study
of the pion field suggests
  that the only solution with the requested mirror symmetries is
the one with equal $\lambda$'s. Thus the valley in the energy for fixed
separation corresponds to the
equal values of $\lambda$ as used in Ref.~\cite{HosakaOA91}.
\begin{figure}
\epsfxsize=10cm
\centerline{\epsffile{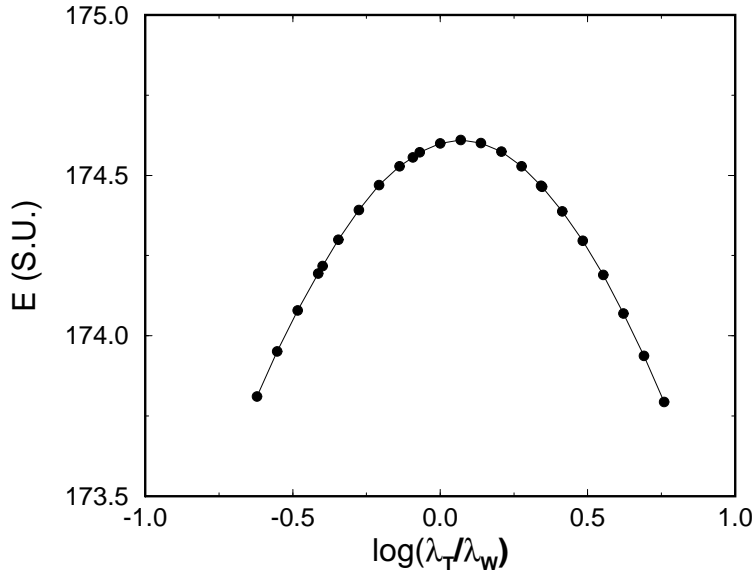}}
\caption{The energy at constant separation in the repulsive channel
as a function of $\log (\lambda_W/\lambda_T)$.
The value of $R$ is $1.3\rm~S.U.$.
\label{fig:Es2}}
\end{figure}
Of much more interest is the repulsive channel, Fig.~\ref{fig:Es2}.
Again a numerical study shows that only for $\lambda_W=\lambda_T$ we
find the correct mirror symmetries for the repulsive channel.
We have chosen to look at a small value of $R$, since only there the energy
depends strongly on $\lambda_W/\lambda_T$, even though the same trend appears
to persist for all $R$. Here we find that the energy is actually a
maximum in the configuration with equal $\lambda$'s (and thus having
the symmetries).
This is not unexpected, since one would like to see that
 in the repulsive channel  only those solutions most repulsive in nature
appear.

\section{Large amplitude collective motion \label{sec:LACM}}

The approach used in selecting collective coordinates is discussed
in detail in Ref.~\cite{KleinWD91}. In this section we shall
discuss the most important ingredients of that work, as well
as the trivial transformation from a Hamiltonian to
a Lagrangian formalism.

\subsection{General Principles}

Given a classical Lagrangian of a general quadratic form in
time derivatives,
\begin{equation}
L = \half \dot \xi^\alpha B_{\alpha\beta}(\xi) \dot \xi_\beta - V(\xi),
\end{equation}
we look for those motions that are approximately decoupled.
A decoupled surface $\Sigma$ is defined by the property that
if we start the  time-evolution from a point on the surface
(i.e., $\xi \in \Sigma$),
with velocity $\dot \xi$ tangential to this surface,
the motion remains on the manifold for all times.
Since a slight perturbation of the Hamiltonian
will destroy such a manifold,
one sees that only very special models will have exact decoupling.

For that reason we look for approximate solutions.
First of all, we select a subset of the exact decoupling
conditions, that allows us to construct an algorithm to calculate
a candidate manifold.
Secondly we construct a decoupling measure, that allows us to gauge
the quality of decoupling. This tells us whether it makes sense to
consider the limited dynamics on the manifold.

Let us start with the first problem, which is the more complicated
of the two. As is argued in Ref.~\cite{KleinWD91}, the mass matrix
$B_{\alpha\beta}$ actually plays the role of metric on the space $q$.
We thus take over the standard notations of general relativity.
In particular $B$ with both indices raises is the inverse mass matrix.
The geometric structure of the problem is especially clear in
in one of the algorithms useful in tackling the decoupling problem,
the local harmonic approximation. This consists of the solution
to the equations (the subscript ``$,\alpha$'' is a shorthand for the partial
derivative $\partial/\partial\xi^\alpha$, ``$;\alpha$'' denotes the
covariant derivative, to be defined below)
\begin{eqnarray}
V_{,\alpha}& =& \sum_i^N \lambda^{(i)} f^{(i)}_\alpha, \label{eq:force}\\
V_{,\alpha;\beta}B^{\beta\gamma}f^{(i)}_\gamma  &=&
(\omega^{(i)})^2 f^{(i)}_\alpha. \label{eq:RPA}
\end{eqnarray}
The first of these equations states that the force is parallel to
an approximate set of tangent vectors to $\Sigma$,
while the second equation is a local harmonic approximation (LHA) to the
motion,
which defines the approximate tangent vectors.
Of great importance is the use of the covariant second derivative of $V$,
\begin{eqnarray}
V_{,\alpha;\beta} &=&
V_{,\alpha,\beta} -V_{,\gamma} \Gamma^\gamma_{\alpha\beta}\nonumber\\
\Gamma^\gamma_{\alpha\beta} &=&
\half B^{\gamma\delta}(B_{\delta \beta,\alpha}+B_{\delta\alpha,\beta}
-B_{\alpha\beta,\delta}).
\end{eqnarray}
We  obtain the ordinary local harmonic approximation
only at extrema of the potential energy ($V_{,\alpha}=0$).
At these points Eq.~(\ref{eq:force}) is automatically satisfied.
This
allows for a path tracking approach, where we bootstrap from a
stationary point.

Let us expand on this point, to clarify our approach.
First note that since Eq.~(\ref{eq:RPA}) is not a symmetric eigenvalue problem,
we have independent left eigenvectors $g^\alpha_{,i}$. With the standard
normalization condition
\begin{equation}
g^\alpha_{,i} f_{,\alpha}^{j} = \delta_i^j,
\end{equation}
these can be interpreted as defining the derivative of the old coordinates
with respect to the new set which exhibits decoupling. The eigenvectors $f$
then define the derivative of new with respect to the old coordinates.
Suppose now that we start from a stable solution, in the case of the
$B=2$ Skyrme model that could be the donut or the hedgehog.
The direction in which we search
are actually the least stable ones (the ones with the lowest
harmonic frequency squared, $(\omega^{(i)})^2)$ in
Eq.~(\ref{eq:RPA}). Suppose that we follow a single  mode,
as we will do in the next section. We then try to track a one-dimensional
manifold by looking for a solution such that the RPA mode $g^\alpha_{,1}$ is
an approximate tangent vector to the path. We thus take a small step
in this direction,
\begin{equation}
\xi^\alpha = \xi^\alpha_0 + \epsilon g^{\alpha}_{,1}.
\end{equation}
In general this point does not lie on the path, i.e.,
it does not exactly satisfy the
force condition (\ref{eq:force}), so that we have to look for a solution
near this point. This is a non-linear problem,
and can be solved in a variety of ways.
In this work we use a very
simple iterative method based on an approximate linearization of
the force equations, as discussed in Appendix \ref{app:soln}.
Once we have located a second point on
this path, we proceed with the same steps sketched above.
For the problem of the Skyrmion, where we need to consider at least
6-dimensional
dynamics to restore all the broken symmetries, we should
actually calculate a many dimensional surface.
We shall not pursue such an approach here, since it leads to significant
topological and numerical complications, and shall concentrate on
a small subset of the collective manifold instead.

Once we have found the coordinate surface, we define coordinates functions
$Q$ decsribing it, and  we introduce
\begin{equation}
\bar V(Q)=V(\xi(Q)),
\end{equation}
and a collective mass $\bar B$, equal to
\begin{equation}
\bar B_{\mu\nu} = g^{\alpha}_{,\mu}B^{\alpha\beta}g^{\beta}_{,\nu}.
\end{equation}
As usual for any linear eigenvalue problem, the scale of the eigenvectors
is undetermined in Eq.~(\ref{eq:RPA}). This freedom corresponds to a
rescaling of the coordinates, which is always possible.
Tensors are also transformed
under such a rescaling. Of particular interest is the mass along the
collective path, which can either be evaluated as
\begin{equation}
B_{11} = g^{\alpha}_{,1}B^{\alpha\beta}g^{\beta}_{,1} ,
\end{equation}
or as
\begin{equation}
\breve B_{11} = \frac{\partial\xi^\alpha}{\partial q}
B_{\alpha\beta} \frac{\partial\xi^\beta}{\partial q}.
\end{equation}
This last definition is based on the exact tangents to the decoupled
manifold. Thus
these definitions are only equal for exact decoupling. The normalized
difference can be used to define a decoupling measure.
If we normalize $g$ such that $B_{ii}=1$,  we find that this
decoupling measure  can be cast in the suggestive form \cite{WaletKD91}
\begin{equation}
D = \sum_{i>2} (dq^i/dq^1)^2.
\end{equation}
Here one can approximate the change in the new coordinates, $dq^i$, by
a finite difference,
\begin{equation}
dq^i \approx f_{,\alpha}^i \delta \xi^\alpha,
\label{eq:defdq}
\end{equation}
which indeed defines a change in the new coordinates. It is obvious
that if all coordinates are independent from the first coordinate
we have exact decoupling.
The steps $dq^1$ can then be used to define a collective coordinate,
\begin{equation}
Q = \sum_{\rm path} dq^1.
\label{eq:defQ}
\end{equation}

\subsection{Redundant coordinates}
One important difference between the current problem and all the
previous ones studied is the appearance of redundant coordinates.
This is easily treated using the local harmonic approximation.
The manifestation of redundant coordinates (that may change from
point to point) is a zero eigenvalue of the mass matrix $B$,
which is thus no longer invertible. But to calculate the covariant derivative
we need this inverse!
The solution is  to only be perform the matrix inversion
 in the subspace with non-zero eigenvalues, the resulting $\Gamma$ correctly
eliminates these non-physical eigenvectors, including their derivatives,
from the LHA equations (\ref{eq:RPA}). Here it is thus of utmost
importance to use covariant derivatives, since the affine connection
contains information about the change of the unphysical modes
with a change of coordinates.
For the present model,
where the extra degrees of freedom can be eliminated explicitly,
 this procedure can easily be shown to give correct results.
We have chosen to use the full set of degrees of freedom
since it provides a sensitive test for the numerical accuracy and
the correctness of our program.

\section{Numerical techniques \label{sec:num}}

{}From the previous sections one may notice that we need the spatial
integration
of the Lagrange density, and the second derivative of the potential energy
as well as the first derivative of the mass. Since the potential energy
depends on the current $l^a_i$, we need up to the second derivative
of $l^a_i$ with respect to $\xi_\alpha$. For the derivative of the mass,
 where the mass itself depends on $l^a_i$ and
$l^a_{,\alpha}$, we need additionally the
derivative of $l_{,\alpha}$. We have chosen to write exact differential
equation for all these quantities, so as not to loose to much accuracy on
the outset. Thus we need to solve, at each point in space, a set of 1269
coupled differential equations. This is solved by first going to the variable
$\tau = \pi/2 \arctan(t)$, and integrating from $-1$ to $1$,
using a highly efficient Runge-Kutta routine
\cite{RK} to solve the initial value problem.

The spatial integral is decomposed into a radial integral and an integral
over the surface of a sphere, which we deform to an ellipsoidal shape by
multiplying
each of the major axes with a constant. The integral over the surface of the
sphere was performed with an efficient non-product formula
(see Ref.~\cite{Stroud71} for general references to such formulas), which
also has at least tetrahedral symmetry. The relevant methods for the
sphere are discussed in
great detail in the Russian mathematical literature \cite{sphere1,%
sphere2,sphere3}.
For the work reported here we use both an 85 point 15th order and a 110 point
17th order formula. These have proven adequate for the task.

The remaining radial integral is transformed to the finite interval $(-1,1)$
by the mapping $\rho=(r-a)/(r+a)$.
We then perform a Gauss-Jacobi integration over $\rho$, using the
weight function $(1-\rho)^2 (1+\rho)^2$. This takes into account
the exact large $r$ behavior of the slowest decaying integrand, which is the
quadratic contribution to the potential energy, as well as the $r^2$ behavior
near the origin.
Judging from the values we found for the baryon number,
this procedure is highly accurate. (We found a value of 2 in
at least six significant digits.) Since the integrands are much
more spread out the
accuracy of the RPA matrices is somewhat less, but we feel that we can
control the numerics.

\section{Results \label{sec:results}}
\subsection{Attractive channel}
\begin{figure}
\epsfxsize=13cm
\centerline{\epsffile{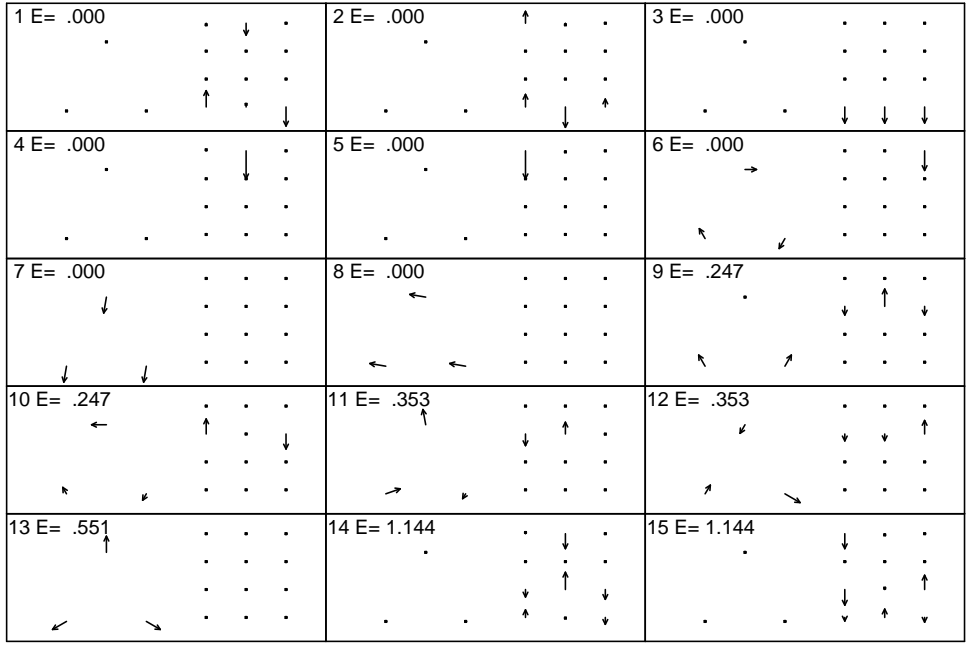}}
\caption{A schematic representations of the eigenvectors of the RPA problem
around the donut. The arrows in each box denote the direction of
the given component of the eigenvector. The triangle on the left gives
the $x,y$ components for each of the three poles. The triangle itself
denotes the position of the poles. The four rows of three points on the
right denote, from
bottom to top, the $z$ components, the time components and the change
in $\lambda$, for (from left to right) poles 1,2,3. The upper line gives
the change in $\theta_i$ (Eq.~(\protect{\ref{eq:deftheta}})).
\label{fig:donut_modes}}
\end{figure}

We start the application of the machinery discussed in the previous sections
by studying the fluctuation around the minimum energy solution, the donut.
In the AM Ansatz this state can be described by a configuration where all three
poles have equal weight $\lambda_i=1$, and the poles themselves are
located in an equilateral triangle, see Fig.~\ref{fig:poles}c.
We first study the harmonic modes around the donut.
The eigenvalues and the eigenfunctions ($g^\alpha_{,\mu}$,
the left eigenvectors of $VB^{-1}$ (Eq.~(\ref{eq:RPA}),
that describe the change of the old
coordinates as we follow each of the modes over an infinitesimal distance),
 are represented graphically in  Fig.~\ref{fig:donut_modes}.

There are eight zero-modes, out of a maximum of nine. These correspond
to the translations in space, rotations in space and rotations in
iso-space. One disappears due to the special
cylindrical symmetry, $2I_3+J_3=0$, of the donut.
Let us discuss the zero-energy part of the spectrum first.
The first and second mode form a pair, describing a rotation around the
$y$ and $x$ axis,
respectively. Since a rotation of the poles around these axes also
generates an overall isorotation, these modes mix with an overall iso-rotation
around the same axis to compensate for this effect. The third mode
describes a translation in the $z$-direction. The fourth and fifth modes
describe iso-rotations around the $y$ and $x$ axis. The sixth mode
describes the effect of the remaining  (iso)rotational
symmetry around the $z$-axis, $I_3-2J_3$. Modes seven and eight are the
translations along the $x$ and $y$ axes (slightly mixed).

In table \ref{tab:1} we list the masses of the zero-modes, when normalized
in such a way that they describe the effect of the standard generators.
\begin{table}
\caption{Masses of zero modes. \label{tab:1}}
\begin{center}
\begin{tabular}{rr|l}
type of mode && mass (S.U.)\\
\hline
translation & $x$ & 67.558\\
translation & $y$ & 67.558\\
translation & $z$ & 76.652\\
rotation & $x$ & 86.964\\
rotation & $y$ & 86.964\\
rotation & $z$ & 130.342\\
isorotation & $x$ & 57.258\\
isorotation & $y$ & 57.258\\
isorotation & $z$ & 32.585\\
off-diagonal &$z$& 65.171
\end{tabular}
\end{center}
\end{table}
One should notice that the mass for translation is not completely isotropic.
This is of course inconsistent with Galilei invariance,
and appears to be incorrect. One can show
that if the AM-donut were the minimum energy solution of the Skyrme equations
this problem would not appear. It is due to the fact that the current
Ansatz is not an exact  stationary solution, and our covariant derivative,
evaluated using the same Ansatz, can not correct for this effect. It
shows some of the limitations of the present approach.

Now let us look at the finite energy modes. The ninth and tenth modes
are the lowest energy ones, and should be the modes we are
looking for. The simplest way to understand the meaning of these modes
is to look at a plot of the pion field in the $xy$ plane. In this plane
the pion
field has no $z$ component, so we can plot it as an arrow in 2D space.
\begin{figure}
\epsfxsize=10cm
\centerline{\epsffile{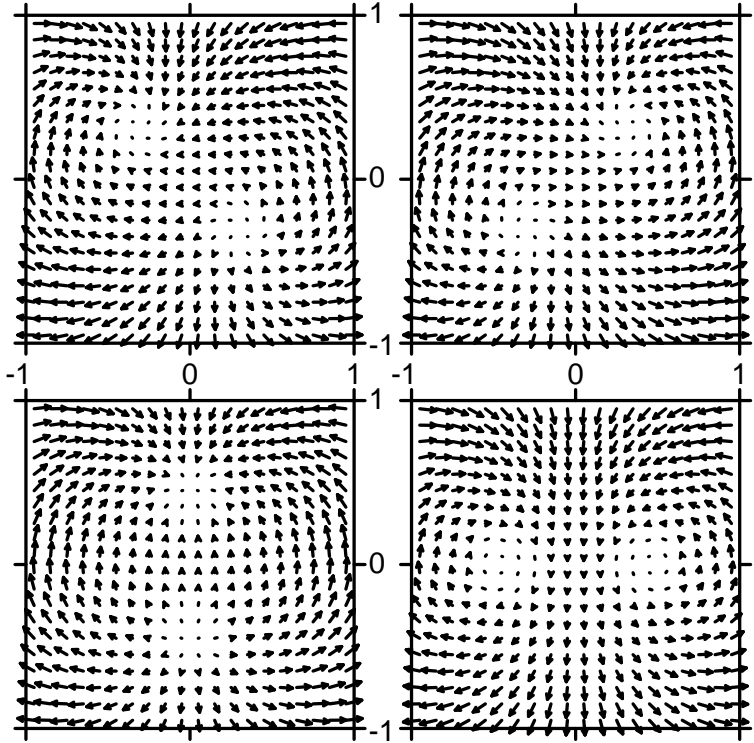}}
\caption{The pion field in the $xy$ plane obtained by following the
ninth (lowest two pictures) mode and the tenth mode (upper two) by a very small
distance. We follow the modes in both positive and negative directions.
\label{fig:arrow_lowest}}
\end{figure}
Such a plot can be found in Fig.~\ref{fig:arrow_lowest}, where we
have followed both these modes by adding and subtracting
a small bit of the relevant mode vectors to the poles specifying the donut.
The idea is that the pion field is only zero at the center of the donut,
and that any perturbation will be clear from a change of the position
of the zeroes. This is clearly the case!
As one can see in the lower two figures, the ninth mode corresponds to
the donut coming apart along either the $x$ or $y$ axis, the axis depending
on the sign of the mode eigenvector. The other mode corresponds to
the system coming apart along an axis making an angle of $45^\circ$ with
the $x$ and $y$ axis. The reason is of course that the donut also has
a reflection symmetry with respect to these axes!
Actually the {\em change} in the pion field is trivial for these modes;
it is a vector field parallel to the $x$ or $y$ axis, respectively.

The eleventh and twelfth modes are somewhat harder to understand.
They look somewhat similar to the previous two modes, if we rotate
the coordinate system by 45 degrees. They are very different, however!
The pion field changes in a complex and much less  symmetric way than
in the case of the
previous modes. In general the trend seems to suggest that these modes
may be those that connect to the hedgehog-hedgehog channel.
The thirteenth mode obviously describes
the breathing mode of the donut. Finally the last two modes describe
a rotations in the time plane. Naively,
as discussed in Appendix \ref{app:symm} this is another class of modes that
are compatible with
the reflection symmetries. These modes are pushed to a high energy due
to a strong mixing with the isorotation ``$C$'' degrees of freedom.
One might be tempted -- as the author was -- to forget about
these global isorotations. This is not even approximately right, however,
since without $C$ these highest two RPA modes become the lowest in energy!
For large distances a similar rotation in the time plane
 on two widely separated hedgehogs corresponds to a zero mode, i.e.,
in that case this mode does not change the energy of the system.

\begin{figure}
\epsfxsize=10cm
\centerline{\epsffile{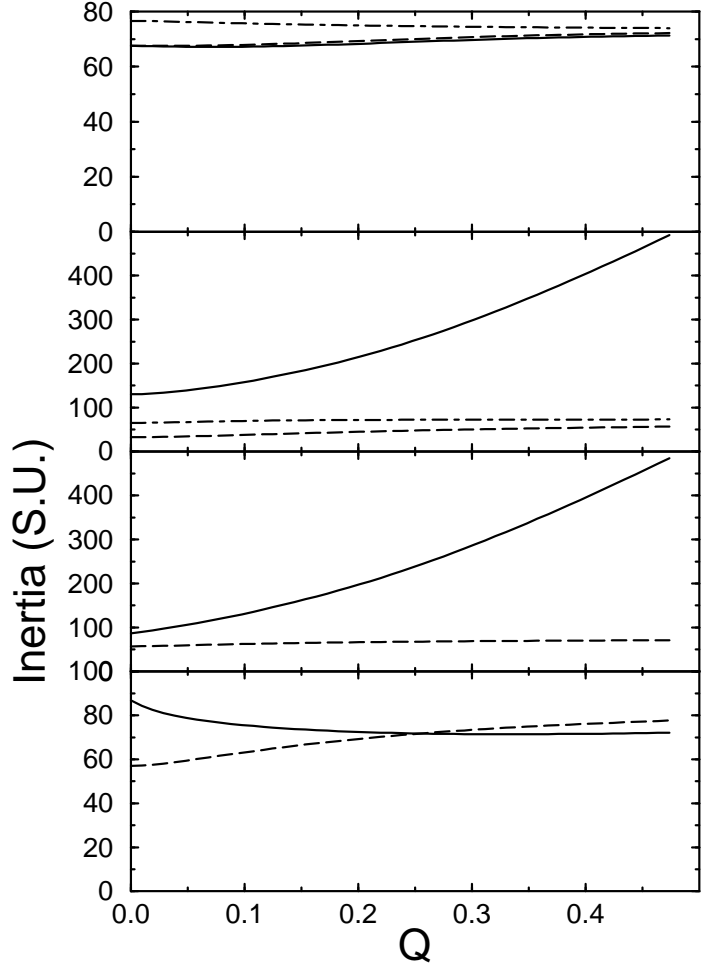}}
\caption{The inertial parameters in the attractive channel, obtained by
following the lowest non-zero mode from the donut.
The lower three panels show the $x$, $y$ and $z$ components of
the (iso)rotational moments of inertia (from bottom to top). Here
solid lines represent rotational moments of inertia, dashed lines isorotational
and dashed-dotted lines the off-diagonal component of the inertia tensor
coupling the two.
The upper panel shows the translational mass.
The solid lines give the $xx$-, the dashed lines the $yy$- and the
dashed-dotted
lines the $zz$-component. Any component of the zero-mode inertia-tensor not
shown is identically zero.
$Q$ is defined in Eq.~(\protect{\ref{eq:defQ}}).
\label{fig:Inertia}}
\end{figure}

\begin{figure}
\epsfxsize=10cm
\centerline{\epsffile{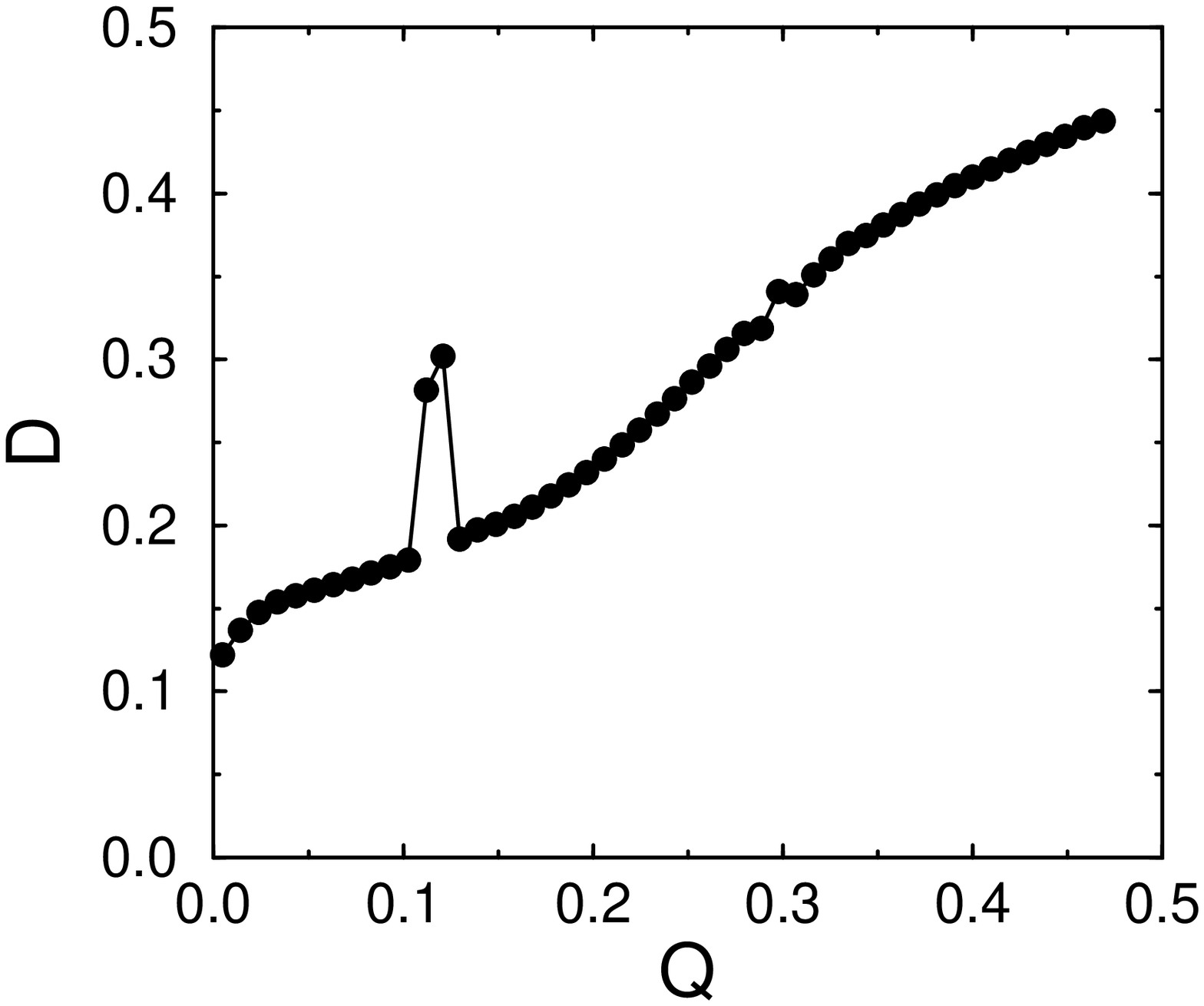}}
\caption{The decoupling measure $D$, Eq.~(\protect{\ref{eq:defD}}),
 in the attractive channel, obtained by
following the lowest mode. $Q$ is defined in Eq.~(\protect{\ref{eq:defQ}}).
The violent jump is due to poor convergence when the RPA frequency
of the mode followed goes through zero.
\label{fig:D}}
\end{figure}

\begin{figure}
\epsfxsize=10cm
\centerline{\epsffile{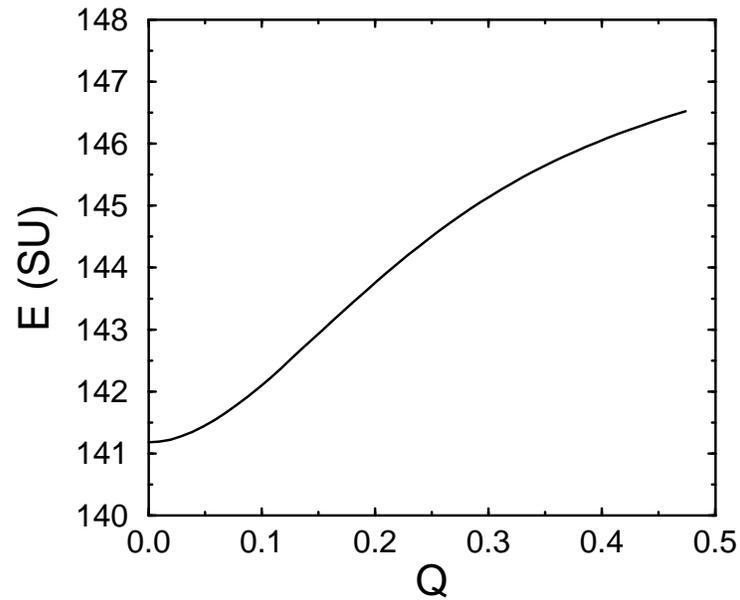}}
\caption{The potential energy for movement in the attractive channel,
 obtained by following the lowest mode.
$Q$ is defined in Eq.~(\protect{\ref{eq:defQ}}).
\label{fig:MV}}
\end{figure}

\begin{figure}
\epsfxsize=10cm
\centerline{\epsffile{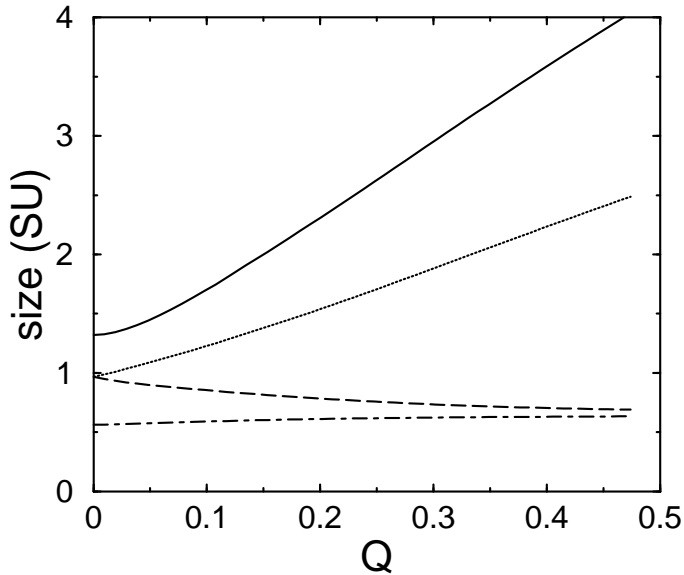}}
\caption[x]{The size of the solution in three directions
in the attractive channel, obtained by
following the lowest mode. $Q$ is defined in Eq.~(\protect{\ref{eq:defQ}}).
The solid line gives $R$, the dotted curve is
\protect{$\sqrt{\langle x^2\rangle}$},
the dashed curve  \protect{$\sqrt{\langle y^2\rangle}$},
 and the dash-dotted curve
 \protect{$\sqrt{\langle x^2\rangle}$}. \label{fig:Size}}
\end{figure}

One of the main goals of the present work was to map out a path in
configuration space where one follows the lowest mode, starting from the
donut. This appears to be rather unproblematic (cf. discussions below),
and there is no major obstacle. Some of the results are presented
in Figs.~\ref{fig:Inertia}--\ref{fig:Size}.

In Fig.~\ref{fig:Inertia} we exhibit the diagonal inertial parameters for
rotations, isorotations and translations (i.e., the mass) as a function
of the coordinate $Q$. These show the expected (and not very exciting)
behavior. The mass is almost constant since the energy is almost constant,
and the breaking of Galilei invariance remains a minor nuisance.
The moment of inertia for rotations increases drastically as $Q$ increases,
for rotations orthogonal to the axis of separation ($x$-axis). Parallel
to this axis it decreases as expected. There is only one off-diagonal component
in the inertia tensor. This is the $z$ component of
the term describing the mixing of the rotational and isorotational momenta.
This term is almost constant. For the hedgehog it has to be there to remove
the symmetry mode from the Hamiltonian. It appears to saturate at a value
of about $73~\rm S.U.$ for large separations.

In Fig.~\ref{fig:D} we show the decoupling parameter $D$, defined in
Eq.~(\ref{eq:defD}). It is small, showing good decoupling. The discontinuity
near $Q$ is 0.1 is due to the fact that there the harmonic energy becomes
zero, and we have mixing with the real zero modes, leading to slightly
anomalous behavior. This is not important for the quality of the solution
outside these points, since the conditions for a valley are local.

In Fig.~\ref{fig:MV} we show the potential energy  $V$ for movement
in the attractive channel. Note that we have defined $Q$ such that the
mass is exactly $1$ so this parameter does not play a role here.

In Fig.~\ref{fig:Size} we show the size of the solution. We also exhibit
the coordinate $R$, defined in Eq.~(\ref{eq:defR}). This can be used
to remap all result in terms of this coordinate. In particular, one
finds that $M(R)=(dQ/dR)^2$, all other quantities being scalars.

\subsection{The hedgehog and the hedgehog-hedgehog channel}
\begin{figure}
\epsfxsize=13cm
\centerline{\epsffile{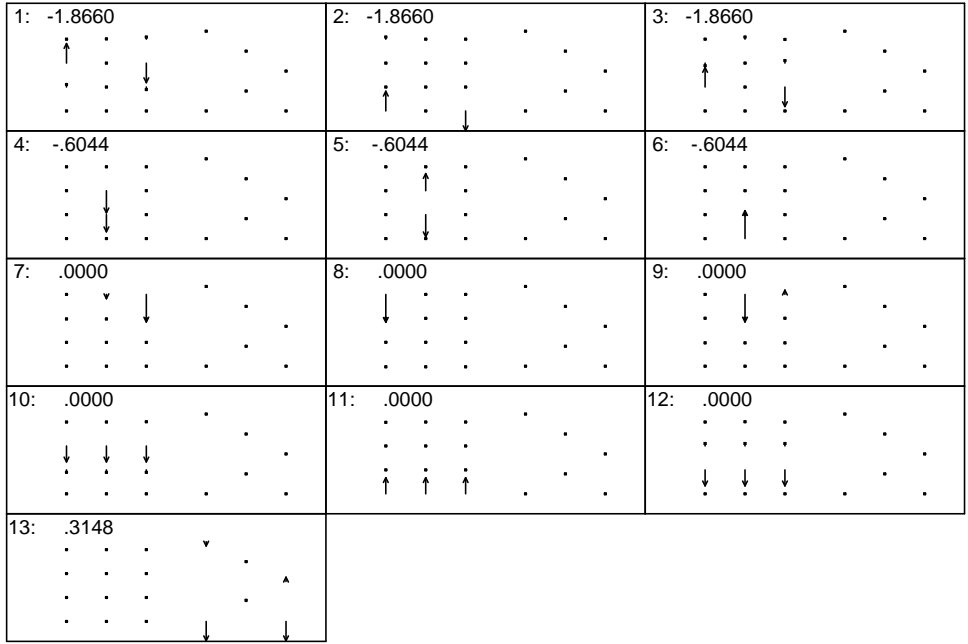}}
\caption{A schematic representations of the eigenvectors of the RPA problem
around the $B=2$ hedgehog.
The arrows in the box denote the direction of change of
the given component of the eigenvector.
The four rows of three points in the left half of the box denote
from bottom to top, the $x$, $y$ and $z$ components, with the upper
line giving the change in $\theta_i$.
The two uneven rows on the right denote the change in $\lambda$ (lower)
and $T$ (upper). We have also indicated the squares of the harmonic
frequencies.
\label{fig:hedge_modes}}
\end{figure}

Let us first study the hedgehog solutions, to see whether there is a
way out of the dilemma that the hedgehog is not connected to the hedgehog
channel. As expected we find two sets of three  negative energy  modes,
one of which corresponds to the hedgehog coming apart by an isorotation
of the individual hedgehogs, and the other to the fission of the
hedgehog into two hedgehogs. Since we have three axes, both sets of modes
are threefold degenerate. In our numerical work following one of the
fissioning modes, we had problems due to a mode that appeared
spurious for the hedgehog (and was thus discarded), which becomes
non-spurious for finite separation. Similar things always occur when
we approach a state where an additional symmetry is realized,
such ad in nuclear physics, if we approach a state of axial
symmetry \cite{RingSchuck}. In these cases the RPA frequency
has a well defined limit as one approaches symmetry, since
$B$ and $V$ (short-hand for the covariant second derivative of $V$)
approach zero at the same rate. Since the harmonic frequency
is the square root of $V/B$, we cannot have that $V$ is finite while $B$
goes to zero. Actually as discussed in great detail in appendix
\ref{app:hedge}, this case is exactly realized in our numerical
calculations! We find it hard to believe
that the Skyrme model itself is so pathological, so we must assume that
this is a problem of the AM Ansatz. That we do not see a way to smoothly
connect the
pole configuration in the hedgehog-hedgehog channel to the hedgehog is
probably closely tied to this problem. It is worrisome and probably
signifies that for strongly interacting Skyrme systems at small distances
we should not take the AM Ansatz to be the final word. For the $NN$ force
this may not be so important, since the Skyrme model itself is suspect at
these distances.

Studying the hedgehog channel is no major problem. The technique we used
is to start with a pole configuration obtained from the ``turning-point
approach'' discussed in Sec.~\ref{sec:SmAM} and iterate until convergence.
At the point thus obtained we follow the relevant mode in both directions
to obtain a picture of this channel. Fortunately the initial guess
was quite close to the actual path.

\begin{figure}
\epsfxsize=10cm
\centerline{\epsffile{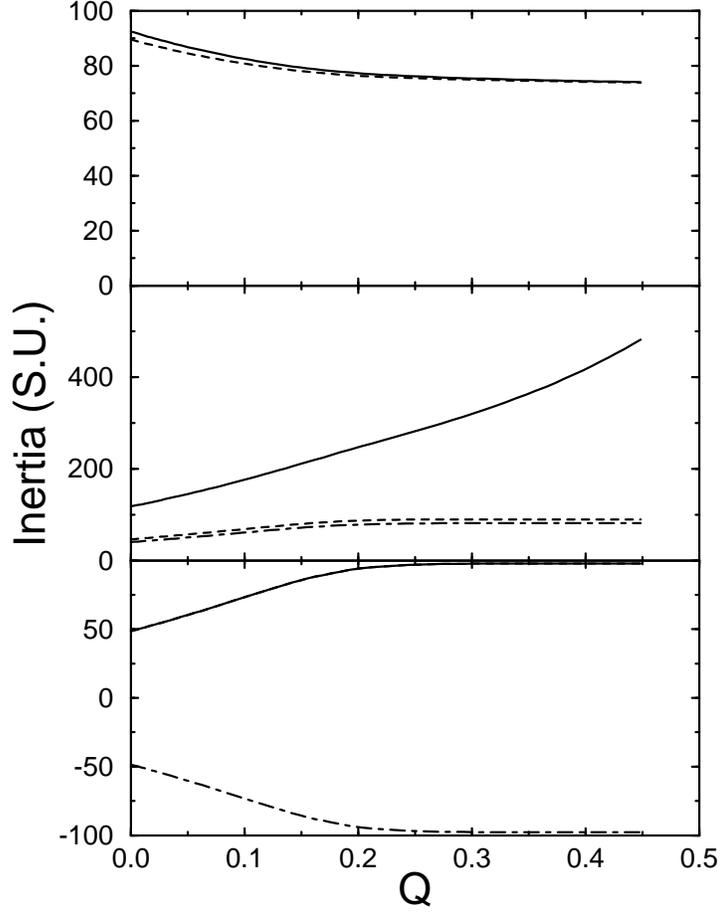}}
\caption{The inertial parameters in the hedgehog-hedgehog channel, obtained by
following the lowest mode with the relevant symmetries.
The lower two panels show the $x$ and the  $y=z$ components of
the (iso)rotational moments of inertia (from bottom to top). Here
solid lines represent rotational moments of inertia, dashed lines isorotational
and dashed-dotted lines the off-diagonal component of the inertia t
ensor coupling the two.
The upper panel shows the translational mass.
The solid lines give the $xx$-, the dashed lines the $yy$- and the
dashed-dotted
lines the $zz$-component. Any component of the zero-mode inertia-tensor not
shown is identically zero.
\label{fig:InertiaHH}}
\end{figure}
\begin{figure}
\epsfxsize=10cm
\centerline{\epsffile{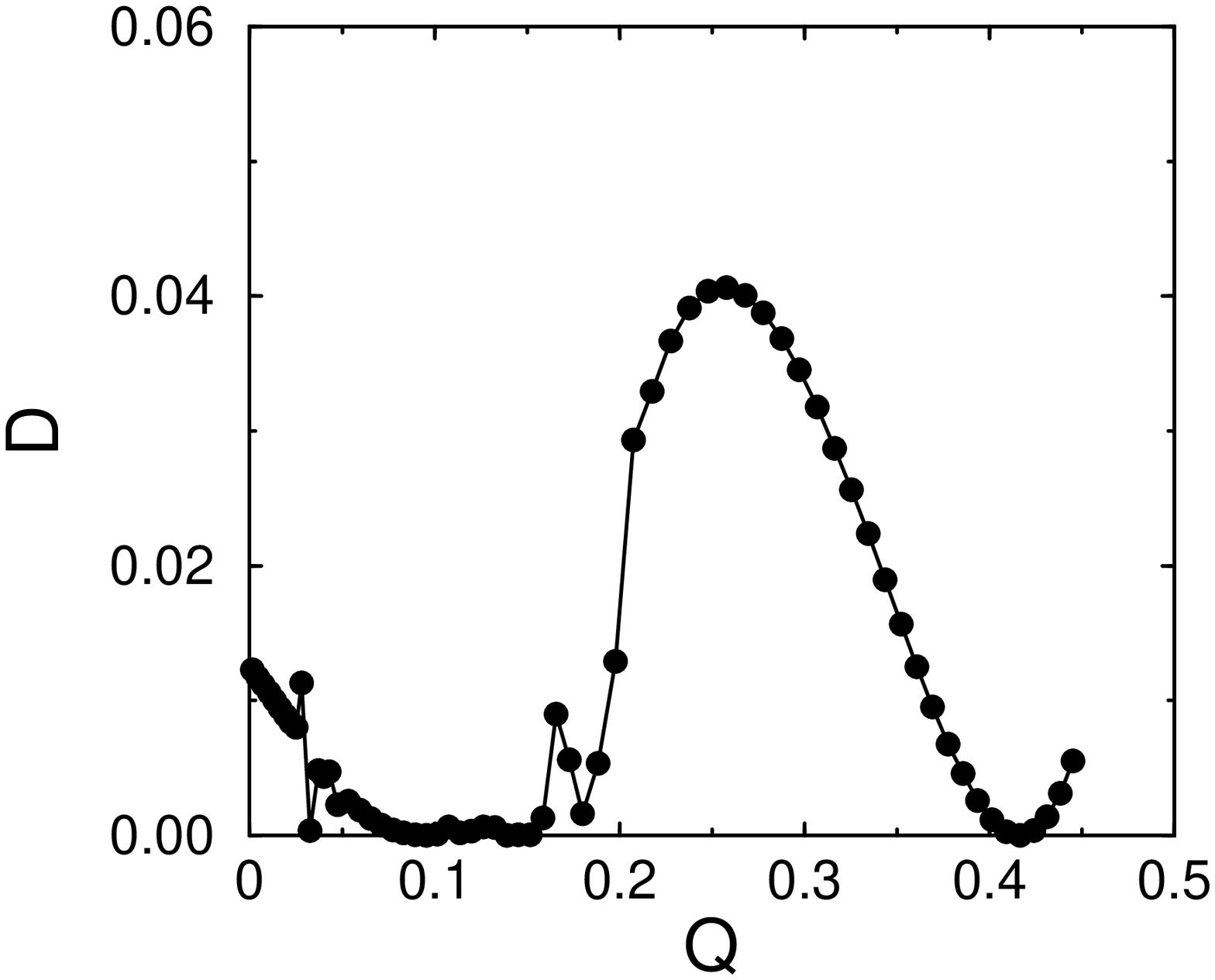}}
\caption{The decoupling measure $D$, Eq.~(\protect{\ref{eq:defD}}),
 in the hedgehog-hedgehog channel, obtained by
following the lowest mode with the correct symmetry.
\label{fig:DHH}}
\end{figure}
\begin{figure}
\epsfxsize=10cm
\centerline{\epsffile{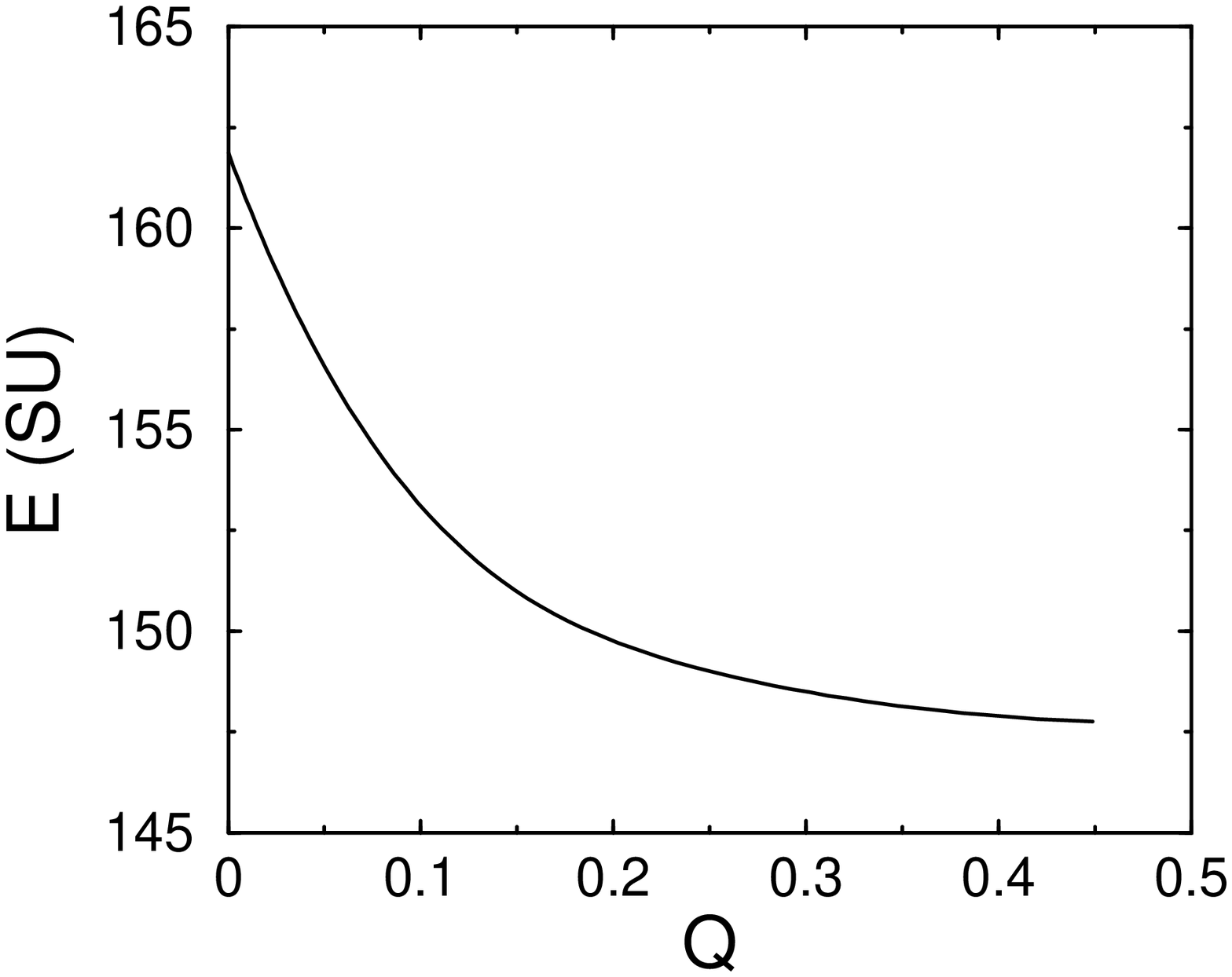}}
\caption{The potential energy for movement in the hedgehog-hedgehog  channel,
 obtained by following the lowest mode.
\label{fig:VHH}}
\end{figure}
\begin{figure}
\epsfxsize=10cm
\centerline{\epsffile{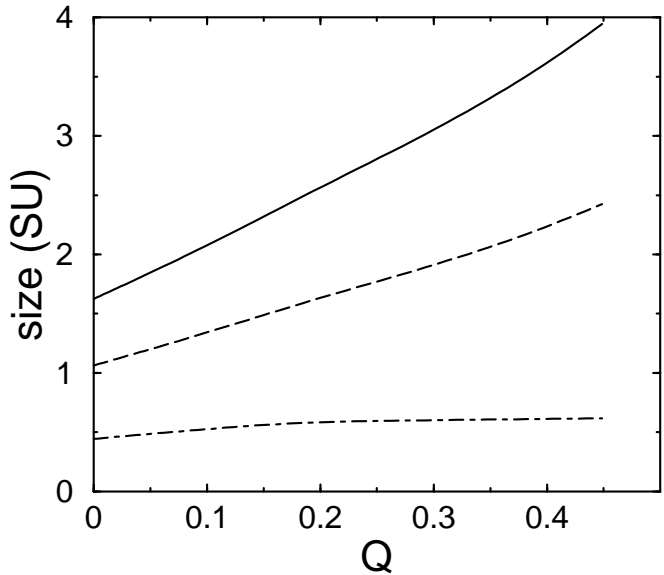}}
\caption[m]{The size of the solution in three directions
in the hedgehog-hedgehog channel, obtained by
following the lowest mode. $Q$ is defined in Eq.~(\protect{\ref{eq:defQ}}).
The solid line gives $R^2$, the dashed line
\protect{$\sqrt{\langle x^2\rangle}$},
the dashed-dotted  line
\protect{$\sqrt{\langle y^2\rangle}=\sqrt{\langle z^2\rangle}$}.
\label{fig:SizeHH}}
\end{figure}

Some of the results are presented in
Figs.~\ref{fig:InertiaHH}--\ref{fig:SizeHH}.
In Fig.~\ref{fig:InertiaHH} we exhibit the inertial parameters for
rotations, isorotations and translations (i.e., the mass) as a function
of the coordinate $Q$. Since the configurations have axial
symmetry around the axis of separation ($x$-axis) the $yy$
and $zz$ components of all these tensors are equal.
The mass is almost constant since the energy is almost constant,
and the breaking of Galilei invariance remains a minor nuisance.
The moment of inertia for rotations increases drastically as $Q$ increases,
for rotations orthogonal to the axis of separation ($x$-axis). Parallel
to this axis it decreases as expected.

There is quite some structure
in the off-diagonal terms. The inertia for ordinary rotations around
the $x$-axis is identical to that for iso-rotations around the same axis,
whereas the mixing term is the opposite of these values. This of course
shows that the states are invariant under the action of simultaneous
rotations and iso-rotations along this axis.
Quantum-mechanically this implies that there can be no dynamical motion along
this axis, and thus no kinetic energy proportional to $(I_x+J_x)^2$.
Furthermore there is an (identical) off-diagonal term connecting the
$y$ and $z$ components of rotations and isorotations.

In Fig.~\ref{fig:DHH} we show the decoupling parameter $D$, defined in
Eq.~(\ref{eq:defD}). This parameter is extremely small everywhere, showing
excellent decoupling, i.e., one coordinate is enough to describe the
dynamics in this channel. The small discontinuity near $Q=.1$ is probably
due to some solutions that are slightly less well converged than at the
neighboring points. This is of course of no relevance, since the
discontinuity in $D$ is so small.

In Fig.~\ref{fig:VHH} we show the potential energy  $V$ for movement
in the hedgehog-hedgehog channel. This shows the expected repulsion in this
channel. Note that the energy of the $B=2$ hedgehog is $219.8\rm~S.U.$,
considerably higher than what we have reached in the current calculation.

In Fig.~\ref{fig:SizeHH} we show the size of the solution. We also exhibit
the coordinate $R$, defined in \ref{eq:defR}. One can see that the
transverse size of the solution shrinks as $R$ decreases.

\subsection{The repulsive channel}

\begin{figure}
\epsfxsize=10cm
\centerline{\epsffile{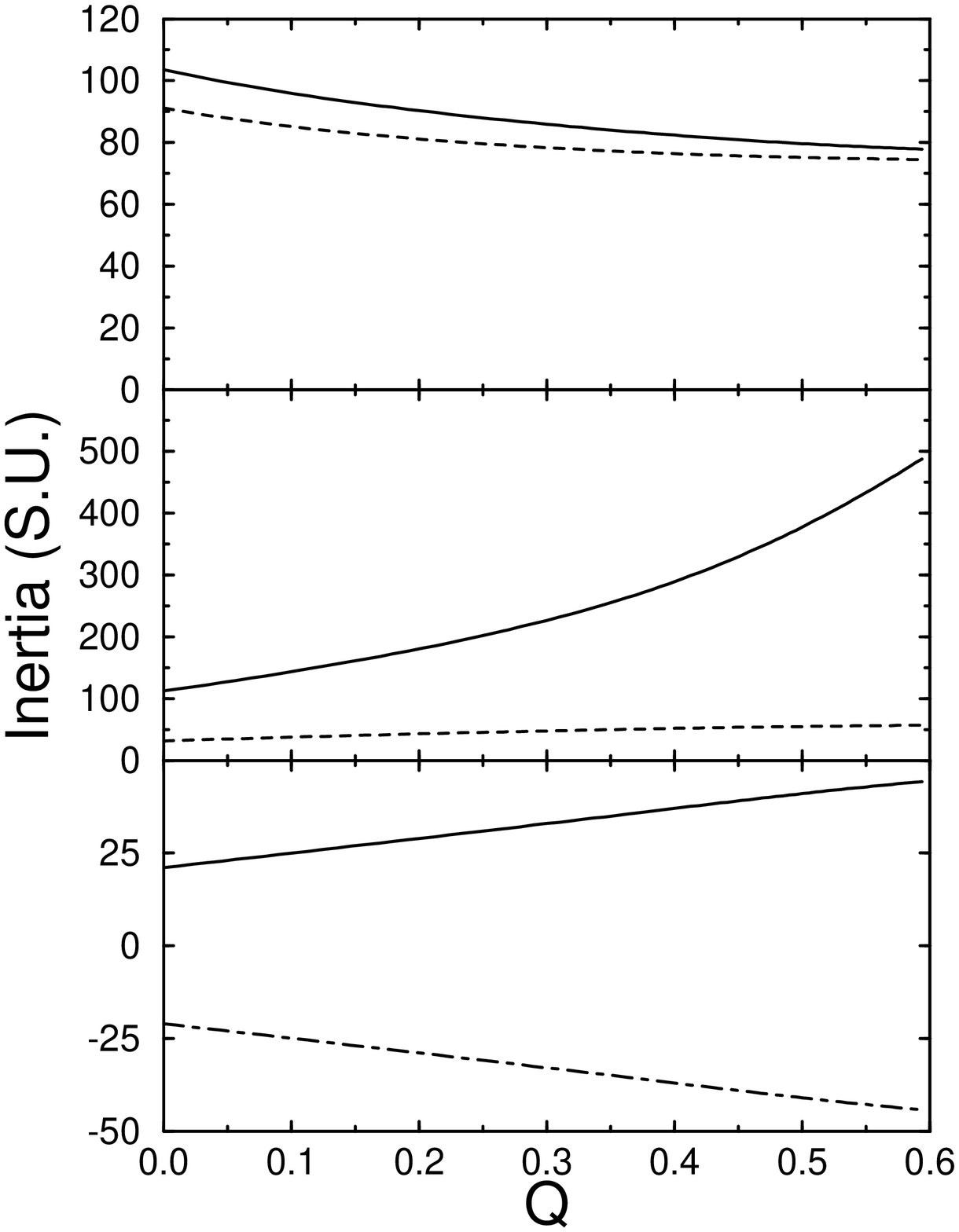}}
\caption{The inertial parameters in the repulsive channel, obtained by
following the lowest mode.
The lower two panels show the $x$ and the  $y=z$ components of
the (iso)rotational moments of inertia (from bottom to top). Here
solid lines represent rotational moments of inertia, dashed lines isorotational
and dashed-dotted lines the off-diagonal component of the inertia tensor
coupling the two.
The upper panel shows the translational mass.
The solid lines give the $xx$-, the dashed lines the $yy$- and the
dashed-dotted
lines the $zz$-component. Any component of the zero-mode inertia-tensor not
shows is identically zero.
\label{fig:Inertiarep}}
\end{figure}

A study of the repulsive channel appears to be more challenging.
The first problem is to find the correct mode to start from.
In order to find this solution we vary the parameter $D$, the size
of the triangle, Eq.~(\ref{eq:tripar}), for a given
value of the parameter $\lambda_2=\lambda_W=\lambda_T$. For a suitably large
value of $\lambda$ (i.e., large separation) we then look for those
values of $D$ such that the valley conditions are satisfied. The problem
arises that there are at least two  solutions that look like they could evolve
to a product Ansatz solution for large separation. We have followed
both of them and rejected one which appears to evolve to a hedgehog that
is too large. A further problem is that in this case the relevant mode
is always the highest in RPA frequency. Since there is a mode the preserves
the symmetry lower in the RPA spectrum, this violates the adiabatic
assumption, and leads to instabilities. The other candidate appears to
describe a solution that is better behaved. The hedgehogs seem a bit small,
but we cannot follow to arbitrarily large separation.

The inertia parameters, Fig.~\ref{fig:Inertiarep},
show similar behavior as in the previous case.
The inertia for ordinary rotations around
the $x$-axis is identical to that for iso-rotations around the same axis,
whereas the mixing term is the opposite of these values. Like in the
hedgehog-hedgehog channel
this implies the absence of terms of the form $(I_x+J_x)^2$ in the Hamiltonian.
There are no further off-diagonal terms.

\begin{figure}
\epsfxsize=10cm
\centerline{\epsffile{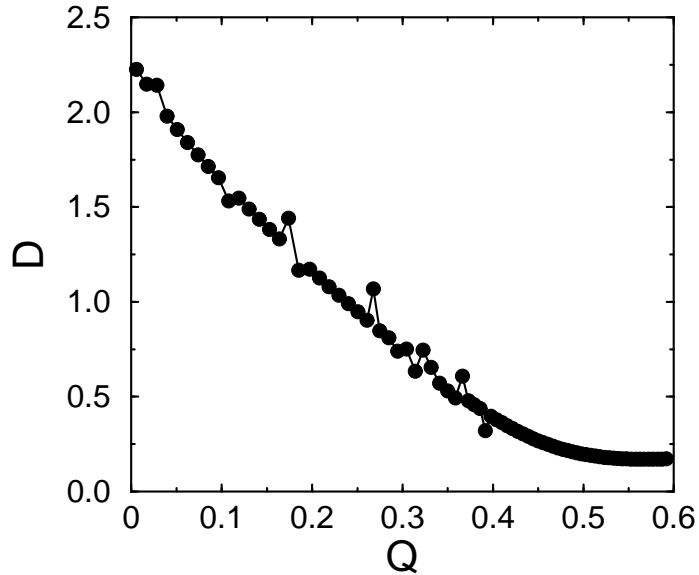}}
\caption{The decoupling measure $D$, Eq.~(\protect{\ref{eq:defD}}),
 in the repulsive channel, obtained by
following the lowest mode with the correct symmetry.
\label{fig:Drep}}
\end{figure}
In Fig.~\ref{fig:Drep} we show the decoupling parameter $D$.
The decoupling measure is large, especially so at small separations.
This indicates that a one-dimensional approach to the repulsive channel is not
sufficient. There are two modes that contribute significantly to $D$. The
manifold of these three modes corresponds exactly to the
parametrization of Eq.~(\ref{eq:tripar}). The one mode that preserves the
reflection symmetries is always higher than the mode we follow, and the one
that
breaks the symmetry considerably lower in the spectrum (in most cases
very close to zero, showing that the energy is almost flat in this
direction), which has to do
with being at a maximum of the energy (cf.~Fig.~\ref{fig:Es2}). We should
at least include all three these coordinates to get a reasonable result
for a decoupled manifold, and thus for the collective Hamiltonian.
The oscillations in $D$ are due to problems to bring the solutions
to convergence.
For such a large decoupling measure, every minor flaw is amplified
significantly.

\begin{figure}
\epsfxsize=10cm
\centerline{\epsffile{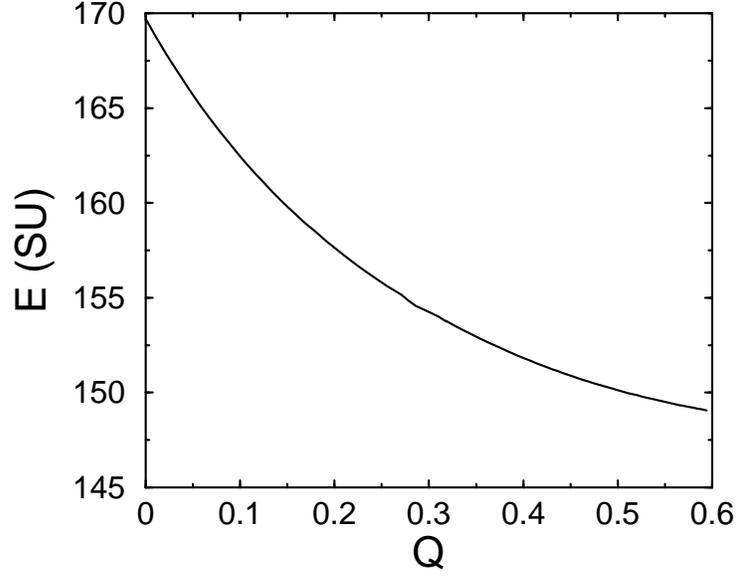}}
\caption{The potential energy for movement in the repulsive channel,
 obtained by following the lowest mode.
$Q$ is defined in Eq.~(\protect{\ref{eq:defQ}})
\label{fig:Vrep}}
\end{figure}
In Fig.~\ref{fig:Vrep} we show the potential energy  $V$ for movement
in the repulsive  channel.  The repulsion is indeed even stronger than in
the hedgehog-hedgehog channel.

\begin{figure}
\epsfxsize=10cm
\centerline{\epsffile{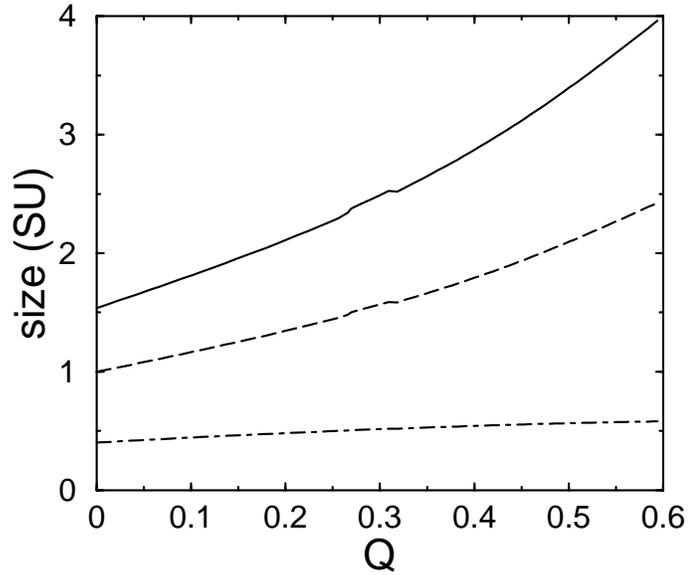}}
\caption[n]{The size of the solution in three directions
in the repulsive channel, obtained by
following the lowest mode. $Q$ is defined in Eq.~(\protect{\ref{eq:defQ}}).
The solid line gives $R$, the dashed line \protect{$\sqrt{\langle
x^2\rangle}$},
the dashed-dotted line
\protect{$\sqrt{\langle y^2\rangle}=\sqrt{\langle z^2\rangle}$}.
\label{fig:Sizerep}}
\end{figure}

\subsection{$NN$ projection}
As discussed in great detail in Ref.~\cite{OBBA87} and Refs.~\cite{WaletAH92,%
WaletAmado93}, one needs at least the three channels discussed here
to project out a nucleon-nucleon force from the classical
Skyrmion-Skyrmion interaction. The standard trick is to use the
parametrization
\begin{equation}
V(\vec R) = v_1(R)+v_2(R)W+v_3(R)Z,
\end{equation}
where $W$ and $Z$ take the values $-1,-2$, $-1,4$ and $3,0$ in the
attractive, repulsive and hedgehog-hedgehog channels, respectively.
One then associates quantum operators with $W$ and $Z$, which allows for
the evaluation of a $NN$ projection of $V$ and by using the
Born-Oppenheimer approximation, the $NN$-force \cite{WaletAmado93}.
Of course this requires the identification of the coordinate $R$ along
the individual paths, which is probably incorrect. A more correct way would
be to find the orthogonal geodesic lines that couple the points $R$ on
one path to the points $R'$ on another, but this appears to be a very
challenging task (if not outright impossible). We shall first convert
the potentials presented previously to functions of $R$.
\begin{figure}
\epsfxsize=10cm
\centerline{\epsffile{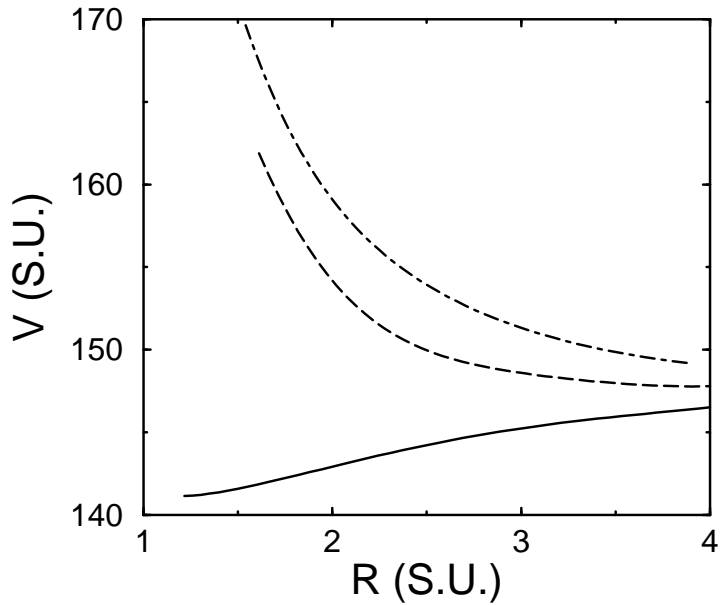}}
\caption{The potential energy
as a function of $R$. The solid line gives the result for the attractive
channel, dashed for the hedgehog-hedgehog channel, and dash-dotted for the
repulsive channel.
\label{fig:EMR}}
\end{figure}
In Fig.~\ref{fig:EMR} these functions are shown. Of some interest is the size
of
the potential energy in the hedgehog channel, which is much lower than in the
product Ansatz.

\begin{figure}
\epsfxsize=10cm
\centerline{\epsffile{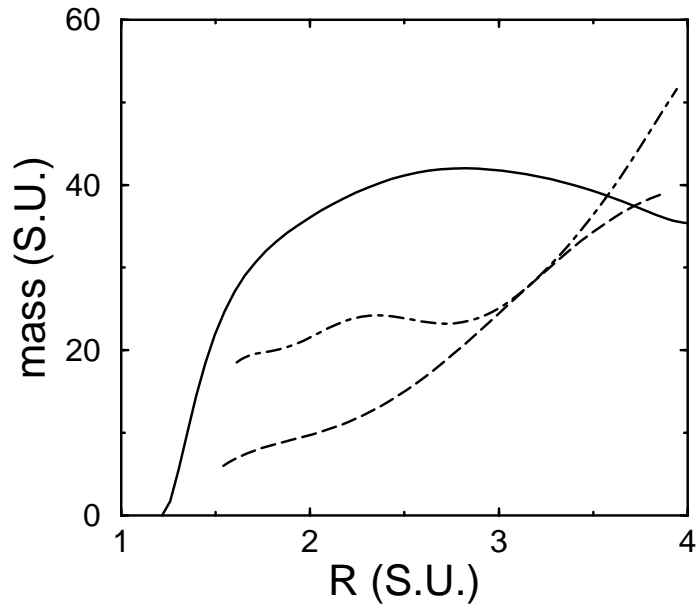}}
\caption{The mass
as a function of $R$. The solid line gives the result for the attractive
channel, dashed for the hedgehog-hedgehog channel, and dash-dotted for the
repulsive channel.
\label{fig:Mtens}}
\end{figure}

The  mass in each channel is $(\frac{dR}{dQ})^2$. In Fig.~\ref{fig:Mtens} we
give
 In the attractive and hedgehog channels this is a smooth function,
which fact is of course closely related to the quality of decoupling.
For large $R$ the mass approaches the reduced mass for two hedgehogs,
$36~\rm S.U.$. In the
repulsive channel the behavior of the mass is surprising, but there decoupling
is poor as well.
One thus finds that the mass to be used in calculations is channel
dependent, in a way similar to the potential energy.

\section{Conclusions\label{sec:disc}}

It appears that the application of LACM techniques to the Skyrme
model is a promising path in the study of the $NN$ force. In this paper we
have applied these techniques to the Atiyah-Manton ansatz, and obtained several
interesting results.  An analysis of the three standard ``paths with
reflection symmetries'' yielded a wealth of information. One of the
important aspects is that the results for the repulsive channel may not be
reliable, since decoupling is poor. This should have an impact on
the calculation of the nucleon-nucleon force from the Skyrme model, but
we have not yet considered this aspect in the current paper. On the
downside it appears that the AM ansatz has some
undesirable problems (such as the lack of a consistent description of
fluctuations around the $B=2$ hedgehog). It would therefore be preferable
to start from an Ansatz that is more fundamental than the add-hoc form of
the AM Ansatz.

One such route would be to apply the techniques of this paper
to the Skyrme model defined on a grid, where the values of $U$ at the
grid points are the relevant dynamical variables. This would also allow
for the addition of a (very physical) finite pion-mass term to the Lagrangian.
The problem with such an approach is that, instead of a few discrete modes, one
 obtains a discretized approximation to bands of states (as in a solid).
Only the band-heads are relevant for the discussion of LACM. The identification
of the different bands is a complicated and highly non-trivial task.
It may actually only be feasible for the channels discussed in this paper,
where symmetries and their representations may be of some help
in classifying the different phonon branches. Still one would
like to be able to connect the different channels, through states without
the additional symmetry, in order to obtain even more information about
the geodesic structure of coordinate space (for a pedestrian: about
the coordinate dependence of the kinetic energy).

Obviously we have not addressed in any detail the structure of the
quantum Hamiltonian, and neither have we used all information derivable
from the present calculation in the construction of the Hamiltonian.
In Ref.~\cite{OBBA87} a boson model was used to analyze the finite $N_c$
effective Hamiltonian for the spin and isospin rotational modes of the
hedgehog. There one concentrated on the leading terms in the large $N_c$
limit, which happens to be the potential energy. In order to make use of
the results of the present paper one should extend the
work of \cite{OBBA87} to include information about the kinetic energy.
One should then calculate the sub-leading terms in the large-$N_c$
limit and match those on the ones calculated in the present work.
Such an investigation is currently underway. Due to the complexity of
this task, this will be described in a separate paper.

\section*{Acknowledgments}
The author wishes to acknowledge Dr. R.D. Amado and Dr. A. Klein
for useful discussions, and Dr. R. Cools for information about
non-product formulas for integration, and sending us a copy
of \cite{Cools}.\\
This work was supported in part by a grant from the
Bundesministerium f\"ur Forschung und Technologie.
The initial stages of the work, at the University of Pennsylvania,
were supported by grants from the United States National
Science Foundation and Department of Energy.

\appendix
\section{Reflections and the product Ansatz \label{app:symm}}
In this appendix we study those solutions of the Skyrme model in the $B=2$
sector that have a simultaneous reflection symmetry in three
orthogonal coordinate planes. To that end we assume that it is enough to
determine what symmetries the product Ansatz has in the limit of
large separations, and then require the same symmetry of
any solution that evolves from it for finite separations.
In this limit
the $U$ fields of the two individual baryon-one hedgehogs commute,
and we can study the symmetries for a single hedgehog,
or the relation of one hedgehog to the other, depending on the
specific reflection plane.

The product
Ansatz is defined by (we rotate one hedgehog by $C$ and the other
by $C^\dagger$, to obtain an initial form with maximal symmetry)
\begin{eqnarray}
U_{\rm PA}({\bf x}|{R}C) & = & U_- U_+ =
U({\bf x}|-{R}/2,C) U({\bf x}|{R}/2,C^\dagger)
\nonumber\\
& = & C[\cos f_- + i\hat x_-\cdot {\bf \tau} \sin f_-] C^\dagger
      C^\dagger[\cos f_+ + i\hat x_+\cdot {\bf \tau} \sin f_+] C.
\end{eqnarray}
Here we use
\begin{eqnarray}
f_\pm &=& f(x_\pm),\;\;{\bf x}_\pm  =  {\bf x} \mp \frac{R}{2} \hat e_1.
\end{eqnarray}
The function $f$ is the chiral profile that describes the hedgehog.

For future use we introduce the rotation matrix $D\equiv D[C]$ associated
with $C$,
\begin{eqnarray}
D_{ij} & = & \frac{1}{2} {\rm tr}[C \tau_i C^\dagger \tau_j].
\label{eq:defD}
\end{eqnarray}
We write
\begin{equation}
\sigma^{(i)}_{kl} = \delta_{kl}(1-2\delta_{ki}).
\end{equation}
for the diagonal matrix with one entry $-1$, which describes
a reflection in a coordinate plane,
We are looking for those forms of the product Ansatz,
where the field
$U=u_4 +i{\bf \tau}\cdot{\bf u}$ after a reflection of the coordinate
system in any of the three coordinate planes is related to itself
by a coordinate independent transformation.

Let us first look at the  $x_{i} \rightarrow -x_{i}$,
$i=2,3$.
These transformations do not interchange the two hedgehogs,
so that we can look for the
symmetries of each hedgehog separately.
Explicitly we find, using Eq.~(\ref{eq:defD}),
\begin{eqnarray}
U_+ & = & \cos f_+ + i(D\hat x_+) \cdot {\bf \tau} \sin f_+
\nonumber \\
& \rightarrow  & \cos f_+ + i(D\sigma^{(i)}\hat x_+) \cdot {\bf \tau} \sin f_+
,\\
U_- & = & \cos f_- + i(D^T\hat x_-) \cdot {\bf \tau} \sin f_-
\nonumber \\
& \rightarrow  & \cos f_- + i(D^T\sigma^{(i)}\hat x_-) \cdot {\bf \tau} \sin
f_-.
\end{eqnarray}
We require that the components of $U_\pm$ transform as
\begin{eqnarray}
u^4_\pm & \rightarrow & u^4_\pm ,\;\;
{\bf u}_\pm  \rightarrow  S^{(i)} \bf{u}_\pm,
\end{eqnarray}
where the matrix $S^{(i)}$ is constant.
Under the remaining transformation $x_1 \rightarrow -x_1$ we do interchange
the two hedgehogs,
\begin{eqnarray}
U_+
& \rightarrow  & \cos f_- + i(D\sigma^{(1)}\hat x_-) \cdot {\bf \tau} \sin f_-,
\\
U_-
& \rightarrow  & \cos f_+ + i(D^T\sigma^{(1)}\hat x_+) \cdot {\bf \tau} \sin
f_+.
\end{eqnarray}
Here we require the intertwining symmetry
\begin{eqnarray}
u^4_\pm & \rightarrow & u^4_\mp, \;\;
{\bf u}_\pm  \rightarrow  S^{(1)} \bf{u}_\mp.
\end{eqnarray}
The symmetries lead to the conditions
\begin{eqnarray}
S^{(i)} D = D \sigma^{(i)},\;\;\;
S^{(i)} D^{T} = D^{T} \sigma^{(i)},\;\;\;i=2,3, \nonumber\\
S^{(1)} D = D^{T} \sigma^{(1)},\;\;\;
S^{(1)} D^{T} = D \sigma^{(1)}.
\end{eqnarray}
These can converted in equations for the $S$'s
\begin{eqnarray}
S^{(i)} = D \sigma^{(i)}D^{T}=
 D^{T} \sigma^{(i)}D,\;\;\;i=2,3, \nonumber\\
S^{(1)}  = D^{T} \sigma^{(1)}D^{T}= D \sigma^{(1)}D.
\label{eq:S}
\end{eqnarray}

Now let us see what these equations, which state
that we have simultaneous symmetry
under all three reflections, have to say about the matrix $D$.
Equating the two forms for $S^{(i)}$ for $i=2,3$ leads to a condition on $D^2$,
\begin{equation}
\sigma^{(i)}D^2 \sigma^{(i)} = D^2.
\end{equation}
The only type of matrix satisfying this
condition for $i=2$ and $3$ simultaneously is a diagonal matrix,
\begin{equation}
D^2 = {\rm diag} (\pm1,\pm1,\pm1),
\end{equation}
with an even number of minus signs.
Thus $D^2$ is either the identity or a rotation around any axis over
an angle $\pi$.
The only form for $C$ allowed is then
$C=\exp(i\frac{{\bf \tau}}{2} \cdot \hat{n}\phi)$ with $\phi=\frac{\pi k}{2}$.
The second of Eqs.~(\ref{eq:S}) shows that $S^{(1)}=S^{(1)T}$.
We find the following three conditions
\begin{eqnarray}
\sin \phi\, \hat{n}_1^2\hat{n}_3 (1- \cos\phi) & = &0, \\
\sin \phi\, \hat{n}_1^2\hat{n}_2 (1- \cos\phi) & = &0, \\
\sin \phi\, \hat{n}_1^2\hat{n}_1 (1- \cos\phi) & = &\hat{n}_1 \sin \phi.
\end{eqnarray}
There clearly exists the trivial solution for $\sin\phi=0$, or$ \phi= k \pi$,
with no restriction on $\hat n$.
Two more interesting  classes of solutions exist as well,
 where $\phi$ can be any multiple of $\pi/2$
\begin{eqnarray}
\hat{n} & = & (1,0,0), \hat{n}  =  (0,\cos\alpha,\sin\alpha).
\end{eqnarray}
Let us now calculate the matrices $S$ for the interesting classes
of solutions.
\begin{enumerate}
\item $\phi=0$: (HH-channel)
\begin{equation}
S^{(i)} = \sigma^{(i)}.
\end{equation}
\item $\hat{n}  =  (1,0,0)$, $\phi=\pm \pi/2$: (repulsive channel)
\begin{equation}
S^{(1)}
        = {\rm diag}(-1,-1,-1),
\end{equation}
\begin{equation}
S^{(2)}
        = {\rm diag}(1,1,-1),
\end{equation}
\begin{equation}
S^{(3)}
        = {\rm diag}(1,-1,1).
\end{equation}
\item $\hat{n}  = (0,\cos\alpha,\sin\alpha)$ ($\sin \phi = 1$ only):
(attractive channel)
\begin{equation}
S^{(1)} = \sigma^{(1)},
\end{equation}
\begin{equation}
S^{(2)} = \left(\begin{array}{lll}
\cos 2\alpha & 2 \cos^2\alpha\sin\alpha & 2 \cos \alpha \sin^2 \alpha \\
2\cos^2\alpha \sin \alpha & 1-2 \cos^4\alpha & -2 \cos^3\alpha \sin \alpha \\
2 \cos \alpha \sin^2 \alpha & -2 \cos^3\alpha \sin \alpha & \cos^4
\alpha+\sin^4\alpha
\end{array} \right),
\end{equation}
\begin{equation}
S^{(3)} = \left(\begin{array}{lll}
-\cos2\alpha & -2 \cos^2\alpha\sin\alpha & -2 \cos \alpha \sin^2 \alpha\\
-2 \cos^2\alpha\sin\alpha & \cos^4\alpha+\sin^4\alpha &
                                        -2 \cos \alpha \sin^3\alpha\\
-2 \cos \alpha \sin^2 \alpha & -2 \cos \alpha \sin^3\alpha &
                        -1 + 4 \cos^2\alpha-2\cos^4\alpha
\end{array} \right),
\end{equation}
\end{enumerate}
The standard reflection symmetries are usually taken to be
the case $\phi=0$ and $\phi=\pi/2$ with $\hat n=(1,0,0)$ and $(0,0,1)$.
For this last unit vector we have
\begin{eqnarray}
S^{(1)} &= &\sigma^{(1)}, \nonumber \\
S^{(2)} &= &{\rm diag}(-1,1,1), \nonumber\\
S^{(3)} &= &{\rm diag}(1,1,-1).
\end{eqnarray}

\section{Analysis of the behavior near the $B=2$ hedgehog \label{app:hedge}}
Here we shall analyze the behavior of the mass matrix and the potential
energy near the $B=2$ hedgehog.
Starting from
\begin{equation}
\rho(x,t) = \frac{1}{x^2+(t-T)^2} +\frac{\lambda}{x^2+t^2} +
\frac{1}{x^2+(t-T)^2},
\label{eq:hedge}
\end{equation}
we shall study the effect of changing
the parameters $\lambda_i$ and $T_i$.
The advantage of such a choice
is that the matrix $A_4$ is Abelian, and for any change the $U$ field
is of hedgehog form,
\begin{equation}
U  = \exp(i\vec{\tau}\cdot\hat{x} \theta(x)).
\end{equation}
It is also not a restriction to study only this limited subset of
modes, since in a numerical calculation we find that they decouple
from the remaining modes.

To calculate the $\theta$-field it is only required to evaluate
a simple integral,
\begin{eqnarray}
\theta(x) &=& x \int_{-\infty}^\infty f(x,t) dt,
\nonumber\\
f(x,t) &=& \partial_{x^2} \ln \rho = \partial_{x^2} \rho / \rho.
\end{eqnarray}
We shall use the subscript $0$ to denote the default
choice of JNR parameters of Eq.~(\ref{eq:hedge}).
If we now analyze the variations of $f$, we find that to lowest
order
\begin{eqnarray}
f&=&f_0 + \frac{1}{\rho_0}\left[
\frac{\delta\lambda_1}{x^2+(t+T)^2}\left\{\frac{1}{x^2+(t+T)^2}-f_0\right\}+
\right . \nonumber\\&& \left.
\frac{\delta\lambda_2}{x^2+t^2}\left\{\frac{1}{x^2+t^2}-f_0\right\}+
\right . \nonumber\\&& \left.
\frac{\delta\lambda_3}{x^2+(t-T)^2}\left\{\frac{1}{x^2+(t-T)^2}-f_0\right\}+
\right . \nonumber\\&& \left.
\frac{2(t+T)\delta
T_1}{(x^2+(t+T)^2)^2}\left\{\frac{2}{x^2+(t+T)^2}-f_0\right\}+
\right . \nonumber\\&& \left.
\frac{2t\delta T_2}{(x^2+t^2)^2}\left\{\frac{2}{x^2+t^2}-f_0\right\}+
\right . \nonumber\\&& \left.
\frac{2(t-T)\delta
T_3}{(x^2+(t-T)^2)^2}\left\{\frac{2}{x^2+(t-T)^2}-f_0\right\}
\right].
\end{eqnarray}
{}From this we can easily read off that there are four out of a total of six
possible changes that leave $\theta$ invariant. These are
\begin{enumerate}
\item
$(\delta\lambda_1,\delta\lambda_2,\delta\lambda_3) = \epsilon(1,\lambda,1)$.
In this case the three first terms inside the curly brackets combine
to $\epsilon \partial\rho_0$, and the three second terms combine to
$-\epsilon \rho_0 f_0 = -\epsilon \partial\rho_0$. These two
contributions cancel. Of course this is the simultaneous rescaling of
all three $\lambda$'s, which is a zero mode for all pole configurations.
\item
$(\delta\lambda_1,\delta\lambda_2,\delta\lambda_3) = \epsilon(1,0,-1)$.
The reason here is time reversal invariance; thus
\[
\int\frac{1}{\rho_0} \frac{1}{x^2+(t+T)^2}
\left\{\frac{1}{x^2+(t+T)^2}-f_0\right\}dt
\]
is invariant under the interchange $T \rightarrow -T$, and the two
contributions cancel.
\item
$(\delta T_1,\delta T_2,\delta T_3) = \epsilon(0,1,0)$.
The integrand is odd under the interchange $t\rightarrow -t$, and
the integral vanishes.
\item
$(\delta T_1,\delta T_2,\delta T_3) = \epsilon(1,0,1)$.
The time-odd parts of the integrand give 0, and the two time-even parts
cancel.
\end{enumerate}
Of course the modes listed as $3$ and $4$ can be combined to the other trivial
zero, a shift in the time-origin, which exists for all possible pole
configurations.
For the case of the hedgehog
we are thus left with two {\em additional} unphysical modes, which are
due to some residual gauge invariance \cite{AtiyahManton89},
 which is unbroken only for
the particular choice of poles discussed here.

The mass matrix, which can be directly related to the first derivative
of $U$ w.r.t. the parameters must have zero eigenvalues in this
space. It is  not so obvious, and we shall argue also not true, that
$V$ has zero eigenvalues in the same subspace. If this were true we would
feel happy discarding this part of the space, since no dynamical information
is contained in these modes.

To study this problem we look at both $B$ and $V$ in the basis
(components, in order, $\delta\lambda_1$, $\delta\lambda_2$, $\delta\lambda_3$,
$\delta T_1$, $\delta T_2$, $\delta T_3$)
\begin{eqnarray}
e_1 & = & \frac{1}{\sqrt{1+2/\lambda^2}}(1/\lambda,1,1/\lambda,0,0,0),
\nonumber\\
e_2 & = & \frac{1}{\sqrt{3}}(0,0,0,1,1,1),
\nonumber\\
e_3 & = & \frac{1}{\sqrt{2}}(1,0,-1,0,0,0),
\nonumber\\
e_4 & = & \frac{1}{\sqrt{2+4/\lambda^2}}(1,-2/\lambda,1,0,0,0),
\nonumber\\
e_5 & = & \frac{1}{\sqrt{2}}(0,0,0,1,0,-1),
\nonumber\\
e_6 & = & \frac{1}{\sqrt{4}}(0,0,0,1,-2,1).
\end{eqnarray}
In this basis $B$ and $V$ have the six-by-six matrix form
\begin{equation}
B = \left(
    \begin{array}{rrrrrr}
        0 & 0 & 0 & 0 & 0 & 0 \\
        0 & 0 & 0 & 0 & 0 & 0 \\
        0 & 0 & 0 & 0 & 0 & 0 \\
        0 & 0 & 0 & 77.593 & 17.385 & 0 \\
        0 & 0 & 0 & 17.385 & 3.895 & 0 \\
        0 & 0 & 0 & 0 & 0 & 0
    \end{array}
    \right),
\end{equation}
\begin{equation}
V = \left(
    \begin{array}{rrrrrr}
        0 & 0 & 0 & 0 & 0 & 0 \\
        0 & 0 & 0 & 0 & 0 & 0 \\
        0 & 0 & 1.175 & 0 & 0 & .5474 \\
        0 & 0 & 0 & 24.176 & 5.426 & 0 \\
        0 & 0 & 0 & 5.426 & 1.212 & 0 \\
        0 & 0 & .5474 & 0 & 0 & .2464
    \end{array}
    \right).
\end{equation}
If one compares these two matrices, the problem is obvious: the two-by-two
block spanned by the third and sixth vector is zero for $B$ (as we had argued
before), but it is non-zero for $V$. Actually the $V$ block has one
zero eigenvalue, but there is one eigenvector with infinite harmonic
frequency ($\omega^2=V/B$)! In the numerical calculations we see this
eigenvalue coming down to finite values as we move away from the
hedgehog. We believe that this  pathology has nothing
to do with the Skyrme model, but is caused by the Ansatz.
The other two-by-two block (fourth and fifth components) has an eigenvalue
very close to zero, but the same eigenvector and eigenvalue
appears both in $V$ and $B$.
This reflects the
fact that the chiral profile is, for large $\lambda$ and $T$, approximately
a function of $\lambda/T^2$. This is only troublesome for numerical
calculations, and causes no theoretical problems.
\section{Self consistent solutions in 1D LACM \label{app:soln}}
Suppose we start from a point $\xi^0$ where
\begin{eqnarray}
V_{,\alpha}(\xi_0) & = & \lambda_1 f^{1}_{,\alpha}(\xi_0), \\
V_{,\alpha;\beta}(\xi_0) B^{\beta\gamma}(\xi_0) f^{\mu}_{,\gamma}(\xi_0) &=&
(\omega^{(\mu)}(\xi_0))^2 f^{\mu}_{,\alpha}(\xi_0),
\end{eqnarray}
are simultaneously satisfied. We could also calculate the
left eigenvectors $g^\alpha_{,\mu}$ of the RPA matrix, which we assume to
be normalized as
\begin{equation}
g^\alpha_{,\mu}f^{\mu}_{,\gamma} = \delta^\alpha_\gamma.
\end{equation}
We now crank by adding $g^\alpha_{,1}$ multiplied by a small number
to $\xi_0$ to obtain a new point $\xi_1$ where these equations are not longer
satisfied. More specific,
\begin{eqnarray}
V_{,\alpha}(\xi_1) & = & \sum_{\mu=1}^N \lambda_\mu f^{\mu}_{,\alpha}(\xi_1),
\\
V_{,\alpha;\beta}(\xi_1) B^{\beta\gamma}(\xi_1) f^{\mu}_{,\gamma}(\xi_1) &=&
(\omega^{(\mu)})^2 f^{\mu}_{,\alpha}(\xi_1),
\end{eqnarray}
where we assume that by taking a small
step from a point where the equations are satisfied, we have
$\lambda_1\gg \lambda_\mu$.

(This last condition is only schematic. Since
$\lambda_\mu$ form the components of a vector, we should require that
\begin{equation}
\bar{B}^{11}\lambda_1^2 \gg
\bar{B}^{\mu\mu}\lambda_\mu^2.
\end{equation}
Here we have used the fact that $\bar{B}$ is diagonal.)

Using a linear approximation to the first derivative of $V$,
\begin{equation}
V_{,\alpha}(\xi_1+\Delta \xi) = V_{,\alpha}(\xi_1)+V_{,\alpha;\beta}(\xi_1)
\Delta \xi^\beta,
\end{equation}
and assuming that the RPA eigenvectors do not change (which is not right,
but iterative use of our update algorithm should be able to correct for this),
we find that the choice
\begin{equation}
V_{,\alpha;\beta}(\xi_1) \Delta \xi^\beta = - \sum_{\mu=2}^N \lambda_\mu
f^{\mu}_{,\alpha}(\xi_1)
\label{eq:solve}
\end{equation}
will lead to a point where the desired equations are satisfied.
(We have neglected one set of first order contributions, originating
in the change of the eigenvector $f^{1}$. This corresponds to assuming
that the derivative of the RPA matrix is 0.)
Let us now try to simplify the solution to (\ref{eq:solve}) as much
as possible:
\begin{eqnarray}
V_{,\alpha;\beta} B^{\beta\gamma} &=& \sum_{\mu} f^\mu_{,\alpha}
(\omega^{(\mu)})^2 g^{\gamma}_{,\mu},\\
B_{\alpha\beta} (V_{,;}^{-1})^{\beta\gamma} & = & \sum_{\mu} f^\mu_{,\alpha}
(\omega^{(\mu)})^{-2} g^{\gamma}_{,\mu},\\
(V_{,;}^{-1})^{\beta\gamma} & = & \sum_{\mu} B^{\beta\alpha} f^\mu_{,\alpha}
(\omega^{(\mu)})^{-2} g^{\gamma}_{,\mu} \nonumber\\
 & = & \sum_{\mu} \bar{B}^{\nu\mu} g^\alpha_{,\nu} (\omega^{(\mu)})^{-2}
g^{\gamma}_{,\mu} \nonumber\\
& = &  g^\alpha_{,\nu} (\bar{V}^{-1})^{\nu\mu} g^{\gamma}_{,\mu} \nonumber\\
& = &  \sum_{\mu}g^\alpha_{,\mu} (\bar{V}_{\mu\mu})^{-1}  g^{\gamma}_{,\mu} .
\end{eqnarray}
Here we have used the fact that in practice $\bar{V}$ is diagonal (in general
we can always make a choice of basis such that this happens).
If we now use $\lambda^{(\mu)} = V_{,\alpha}g^\alpha_{,\mu}$,
we have
\begin{eqnarray}
\Delta \xi^\alpha &=& - \sum_{\mu=2}^N
\sum_{\mu'}g^\alpha_{,\mu'} (\bar{V}_{\mu'\mu'})^{-1}
 g^{\gamma}_{,\mu'} \lambda_\mu f^{\mu}_{,\gamma}
\nonumber\\
&=&- \sum_{\mu=2}^N g^\alpha_{,\mu} (\bar{V}_{\mu\mu})^{-1}\lambda_\mu
\nonumber\\
&=&- \sum_{\mu=2}^N g^\alpha_{,\mu} (\bar{V}_{\mu\mu})^{-1}g^\gamma_{,\mu}
V_{,\gamma}.
\end{eqnarray}
\end{document}